\documentclass[a4paper,12pt]{article}
\usepackage{amsmath,amssymb}
\usepackage[pdftex]{graphicx}
\usepackage[pdftex]{hyperref}
\usepackage{tikz}
\hypersetup{colorlinks=true,linkcolor=blue,urlcolor=blue,filecolor=blue,citecolor=blue}
\usepackage{epsf}
\usepackage{cite}
\usepackage{fancyhdr}
\usepackage{enumerate} % \begin{enumerate}[(i)] etc.

\makeatletter

\g@addto@macro\bfseries{\boldmath}

\def\refchecklabelfontsize{\fontsize{5pt}{5pt}\selectfont}
\let\mark@size=\refchecklabelfontsize

\def\half{{\frac{1}{2}}}

\def\p{\partial}

\def\unit{{1\kern-.65ex {\rm l}}}
\def\1{{1\kern-.65ex {\rm l}}}

\def\Tr{\mathop{\mathrm{Tr}}\nolimits}

\def\ket#1{{|{#1}\rangle}}

\def\kett#1{{|{#1}\rangle\!\rangle}}

\def\htilde{{\widetilde{h}}} 
\def\jt{{\widetilde{j}}}

\def\Gt{{\widetilde{G}}}

\def\Jt{{\widetilde{J}}}
\def\Lt{{\widetilde{L}}}

\def\cN{{\cal N}}
\def\cO{{\cal O}}

\def\bbZ{{\mathbb{Z}}}

\newcount\hour \newcount\minute
\hour=\time \divide \hour by 60
\minute=\time
\def\now{%
\ifnum \hour<13
  \ifnum \hour=0 \advance \hour by 12 \number\hour:\else \number\hour:\fi%
     \ifnum \minute<10 0\fi%
     \number\minute%
\ A.M.%
\else \advance \hour by -12 \number\hour:%
  \ifnum \minute<10 0\fi%
  \number\minute%
  \ P.M.%
\fi%
}

\makeatother

\begin{document}

% format
\baselineskip=18pt  % a la harvmac
\numberwithin{equation}{section}  % make eq labels (sec.num)
%\allowdisplaybreaks  % allow page breaks in displayed eqs

%% page style
%% print date, time and filename 
%\pagestyle{myheadings}
%\markright{{\tt \jobname.tex} -- \today{} \now}
% print more (needs fancyhdr.sty)
%\pagestyle{fancy}
%\lfoot{{\tt \jobname.tex} -- \today{} \now}
%\cfoot{}
%\rfoot{\thepage}
%\renewcommand{\footrulewidth}{0.4pt}
%\renewcommand{\headrulewidth}{0pt}
%
% If want a shorter version in the page headers:
%   \section[short version]{full version}
% or
%   \section[short version]{full version}
%   \sectionmark{very short version}
% short version appears in the ToC and the page header.
% If very short version is specified, appears in the page header.
% http://www.tex.ac.uk/FAQ-runheadtoobig.html
% same for subsection.
%\section[middling version]{verbose version%
%              \sectionmark{terse version}}
%\sectionmark{terse version}

%%%%%%%%%%%%%%%%%%%%%%%%%%%%%%%%%%%%%%%%%%%
%%%        TITLE BEGINS HERE
%%%%%%%%%%%%%%%%%%%%%%%%%%%%%%%%%%%%%%%%%%%

%% ========== title (note version) begins here ==========

%\vspace*{-1cm}
%\begin{center}
% {\Large\bf Title of the Document}\\
% Masaki Shigemori
%\end{center}
%\vspace*{-.5cm}

%% ========== title (note version) ends here ==========

%% ========== title (paper version, a la harvmac) begins here ==========
%
\thispagestyle{empty}

% Report number
\vspace*{-2cm} 
\begin{flushright}
%{\tt arXiv:yymm.nnnn}\\
%QMUL-PH-17-XX\\
YITP-19-61
\\
\end{flushright}

% title, authors, affiliation
\vspace*{2.5cm} 
\begin{center}
 {\LARGE Counting Superstrata}\\
 \vspace*{1.7cm}
 Masaki Shigemori\\
 \vspace*{1.0cm} 
Department of Physics, Nagoya University, 
Furo-cho, Chikusa-ku, Nagoya 464-8602, Japan\\
and\\
Yukawa Institute for Theoretical Physics (YITP), Kyoto University\\
Kitashirakawa Oiwakecho, Sakyo-ku, Kyoto 606-8502 Japan
  \end{center}
\vspace*{1.5cm}

% abstract
\noindent

We count the number of regular supersymmetric solutions in supergravity,
called superstrata, that represent non-linear completion of linear
fluctuations around empty ${\rm AdS}_3\times S^3$.  These solutions
carry the same charges as the D1-D5-P black hole and represent its
microstates.  We estimate the entropy using thermodynamic approximation
and find that it is parametrically smaller than the area-entropy of the
D1-D5-P black hole.  Therefore, these superstrata based on ${\rm
AdS}_3\times S^3$ are not typical microstates of the black hole.  What
are missing in the superstrata based on ${\rm AdS}_3\times S^3$ are
higher and fractional modes in the dual CFT language.  We speculate on
what kind of other configurations to look at as possible realization of
those modes in gravity picture, such as superstrata based on other
geometries, as well as other brane configurations.

\newpage
\setcounter{page}{1} % don't number title page

% ========== title (paper version, a la harvmac) ends here ==========

%%%%%%%%%%%%%%%%%%%%%%%%%%%%%%%%%%%%%%%%%%%
%%%           TITLE ENDS HERE
%%%%%%%%%%%%%%%%%%%%%%%%%%%%%%%%%%%%%%%%%%%

%\tableofcontents
%\printindex

%%%%%%%%%%%%%%%%%%%%%%%%%%%%%%%%%%%%%%%%%%%
%%%        MAIN TEXT BEGINS HERE
%%%%%%%%%%%%%%%%%%%%%%%%%%%%%%%%%%%%%%%%%%%

\section{Introduction and summary}
\label{sec:intro_summary}

The D1-D5 system plays a central role in the string-theory
understanding of microscopic physics of black holes.  This system is
obtained by compactifying type IIB string theory on~$S^1\times M_4$ with
$M_4=T^4$ or K3 and wrapping $N_1$ D1-branes on~$S^1$ and $N_5$
D5-branes on~$S^1\times M_4$.  If we add a third charge, $N_p$ units of
Kaluza-Klein momentum (P) charge along~$S^1$, we have a 1/8-BPS,
3-charge black hole with a finite entropy which can be reproduced by
microstate counting in the brane worldvolume theory
\cite{Strominger:1996sh}.  More generally, we can also add
left-moving angular momentum $J$ and the area entropy of the resulting
1/8-BPS black hole (the BMPV black hole \cite{Breckenridge:1996is}) is
given by\footnote{$J\in\bbZ/2$ in our convention.}
\begin{align}
 S_{\rm BMPV}=2\pi\sqrt{N N_p-J^2},\qquad N\equiv N_1 N_5.\label{S_BMPV}
\end{align}

Counting microstates in the brane worldvolume theory does not give us
much information about their physical nature in the gravity (bulk)
picture.  Motivated by Mathur's fuzzball conjecture~\cite{Mathur:2005zp,
Mathur:2009hf}, a lot of endeavor has been made to construct microstates
of black holes, especially of the D1-D5-P 3-charge black hole in the
form of ``microstate geometries'', namely, smooth horizonless solutions
of classical supergravity.\footnote{Mathur's conjecture {\it per se\/}
does not claim that general microstates are describable within classical
supergravity.  The microstate geometry program is about how far one can
go with classical supergravity. } In this paper, we will restrict
ourselves to BPS microstate geometries, which are in good theoretical
control.
In the 2-charge case where $N_p=0$, the microstates can be realized in
supergravity as the so-called Lunin-Mathur geometries
\cite{Lunin:2001jy, Lunin:2002iz, Taylor:2005db, Kanitscheider:2007wq},
which are parametrized by functions of one variable.  The growth of the
microscopic entropy, $S_{\text{2-chg}}\sim \sqrt{N}$, can be reproduced
by counting Lunin-Mathur geometries \cite{Rychkov:2005ji,
Krishnan:2015vha}, although the 2-charge system has vanishing area
entropy at the classical level.
In the 3-charge case, for which the area entropy is non-vanishing at the
classical level, many families of microstates have been constructed based
on the multi-center solutions \cite{Bena:2006kb, Bena:2010gg,
Heidmann:2017cxt, Bena:2017fvm} (see \cite{Bena:2005va, Berglund:2005vb}
about smooth multi-center solutions) and other methods, such as
solution-generating technique, the matching technique, and BPS equations
\cite{Mathur:2003hj, Lunin:2004uu, Giusto:2004id, Giusto:2004ip,
Giusto:2006zi, Ford:2006yb, Mathur:2011gz, Mathur:2012tj, Lunin:2012gp,
Giusto:2013bda, Giusto:2012yz}.

More recently, based on a linear structure of BPS equations in 6D
supergravity \cite{Bena:2011dd}, a new class of microstate geometries
called superstrata was constructed \cite{Bena:2015bea, Bena:2016agb,
Bena:2016ypk, Bena:2017geu, Bena:2017xbt, Bakhshaei:2018vux,
Ceplak:2018pws, Heidmann:2019zws}.  Superstrata are microstate
geometries of the 3-charge D1-D5-P black hole, parametrized by functions
of three variables, and their CFT duals are explicitly known.
In essence, superstrata are non-linear completion of the linear
excitations around empty AdS$_3\times S^3$, which are sometimes called
``supergravitons''.  In other words, superstrata represent coherent
states of the supergraviton gas with backreaction.

Because superstrata represent the largest known class of microstate
geometries for the 3-charge black hole, it is of natural interest to
count them and compute their entropy.  In this paper, we carry out such
computation and find an explicit formula for the entropy $S_{\rm
strata}(N,N_p,J)$ for large $N\sim N_p\gg 1$, $|J|=\cO(N)$.  The explicit
functional form of $S_{\rm strata}$ turns out to be quite complicated
because, depending on the values of $N_p$ and $J$, different bosons
condense, which leads to different functional forms of $S_{\rm strata}$.
The explicit formulas are given in 
section \ref{ss:full_pf}.
One interesting regime is the Cardy regime, $N_p\gg N$.\footnote{Note
that we have already taken the large $N,N_p$ limit.  So, this means
$N_p\gg N\ggg 1$.\label{ftnt:Np>>N>>>1}} In particular, for $J=0$, we
find that the entropy behaves as
\begin{align}
 S_{\rm strata}|_{N_p\gg N,J=0}\sim 
N^{1/2}N_p^{1/4}.\label{S_strata_Cardy}
\end{align}
On the other hand, \eqref{S_BMPV} gives
\begin{align}
  S_{\rm BMPV}|_{J=0}
 \sim N^{1/2} N_p^{1/2},
\end{align}
which is parametrically larger than \eqref{S_strata_Cardy}.  Outside the
Cardy regime, the behavior of the superstrata entropy is not simple, but
its parametric growth for $N\sim N_p \sim |J|$ has the following
universal form:
\begin{align}
 S_{\rm strata} \sim N^{3/4} \ll S_{\rm BMPV}\sim N.
 \label{S_strata_estimate}
\end{align}
Therefore, superstrata around AdS$_3\times S^3$, although parametrized
by functions of three variables, have parametrically smaller entropy
than the 3-charge black hole.  This is actually expected, because
superstrata around AdS$_3\times S^3$ involve no higher or fractional
modes that are important for reproducing the black-hole entropy
\cite{Bena:2014qxa, Bena:2015bea, Bena:2016agb}.

The result \eqref{S_strata_estimate} does \emph{not} yet necessarily
mean that supergravity solutions are insufficient for reproducing the
black-hole entropy \eqref{S_BMPV}.  What we counted in
\eqref{S_strata_estimate} are superstrata obtained by putting
supergravitons in empty AdS$_3\times S^3$, but there also exist
superstrata that correspond to putting supergravitons in different
backgrounds.  In particular, in \cite{Bena:2016agb}, superstrata on
$({\rm AdS}_3\times S^3)/\bbZ_k$ backgrounds were constructed, and those
superstrata include some of fractional excitations, which we just said
are important ingredients in order to reproduce the black-hole entropy
in gravity picture.
Therefore, the result \eqref{S_strata_estimate} is better interpreted as
suggesting where the microstate geometry program must go, for it to have
any chance of succeeding in reproducing the black-hole
entropy;\footnote{One could also consider superstrata on backgrounds
with more than non-trivial 3-cycles but, according to the recent work
\cite{Bossard:2019ajg}, they would correspond to microstates of
multi-center black holes, which do not exist everywhere in the moduli
space and thus are not counted by a supersymmetric index.} one must
seriously look into solutions involving higher and fractional
excitations, not only the ones in \cite{Bena:2016agb} but also more
general ones.  We will discuss what kind of configurations to look at in
more detail in section \ref{sec:discussion}.

Our working assumption in computing the entropy for superstrata is that
their geometric quantization will exactly give the Hilbert space of
supergravitons in the AdS$_3\times S^3$ background, with an appropriate
stringy exclusion principle imposed.  This is a very natural and safe
assumption from the proposed holographic dictionary for superstrata
\cite{Bena:2015bea, Bena:2016ypk, Bena:2016agb}, which is almost obvious
from the construction and has passed some non-trivial test
\cite{Giusto:2015dfa}. Under this assumption, we can count superstrata
simply by counting states in the supergraviton Hilbert space, or
equivalently, counting CFT states dual to them.  More precisely, we
compute the relevant partition function and estimate it for large $N\sim
N_p\sim J$ using thermodynamic approximation, obtaining the result
\eqref{S_strata_estimate}.

\bigskip

Let us very briefly recall the structure of the
states of the D1-D5-P system.
In the decoupling limit, the D1-D5 system is dual to a 2-dimensional
boundary CFT called the D1-D5 CFT with central charge $c=6N$, as will be
discussed in more detail later.  We can talk about the microstates of
the D1-D5-P system in the language of this CFT\@.  

\begin{figure}[htb]
 \begin{quote}
 \begin{center}
     \includegraphics[height=5cm]{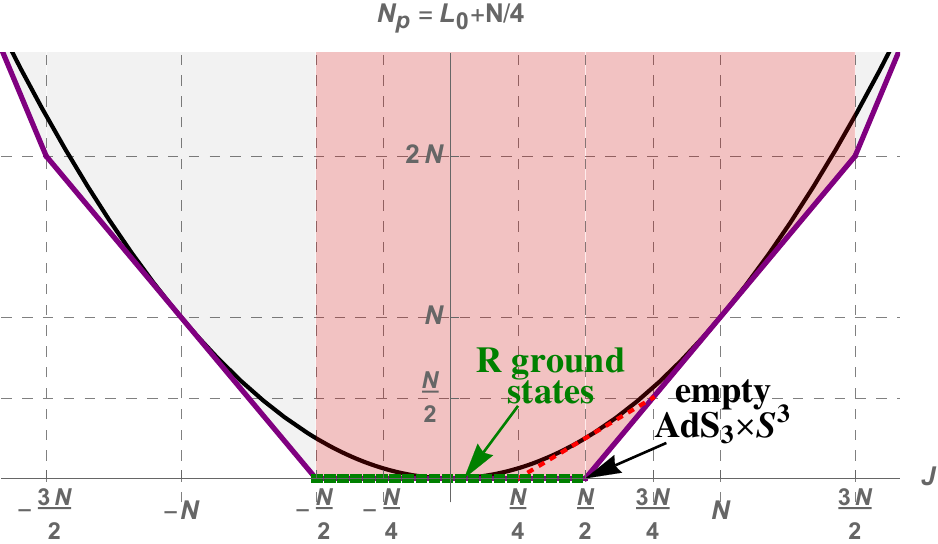}
\caption{\sl The $J$-$N_p$ plane of the 3-charge system in the R sector.
 The states exist only on and above the unitarity bound represented by
 the purple polygon.  The single-center BMPV black hole exists only
 above the parabola $N_p=J^2$ (sometimes called the cosmic censorship
 bound) (black curve).  The empty ${\rm AdS}_3\times S^3$ correspond to
 the point $(J,N_p)=(N/2,0)$.  R ground states live on the green dotted
 line.  The red dotted line represents de Boer's bound (see section
 \ref{sec:discussion}).
  \label{fig:JNp_plane_R}}
 \end{center}
 \end{quote}
\end{figure}

In the Ramond (R) sector of the CFT, on the $J$-$N_p$ plane, states
exist only in the region bounded below by the unitarity bound (the
purple polygon in Figure~\ref{fig:JNp_plane_R}).  Here, $N_p=L_0-{N/ 4}$
and $L_n$ are Virasoro generators.  This is only for the left-moving
sector, but the right-moving sector is similar.
The empty ${\rm AdS}_3\times S^3$ corresponds to the point
$(J,N_p)=({N/ 2},0)$.  We can think of other states as excitation of
this state.
The 2-charge states corresponds to going horizontally to RR ground
states on the interval $J\in [-{N/ 2},{N/ 2}]$, $N_p=0$ (the green
dashed line in Figure~\ref{fig:JNp_plane_R}).
The 3-charge states correspond to going to $N_p>0$.  The 3-charge BMPV
black hole exists only above the parabola $N_p={J^2/ N}$, which is
finitely away from the empty ${\rm AdS}_3\times S^3$ point.
States representing superstrata obtained by exciting supergravitons on
${\rm AdS}_3\times S^3$ exist in the red shaded region in
Figure~\ref{fig:JNp_plane_R}.

\begin{figure}[htb]
 \begin{quote}
 \begin{center}
     \includegraphics[height=5cm]{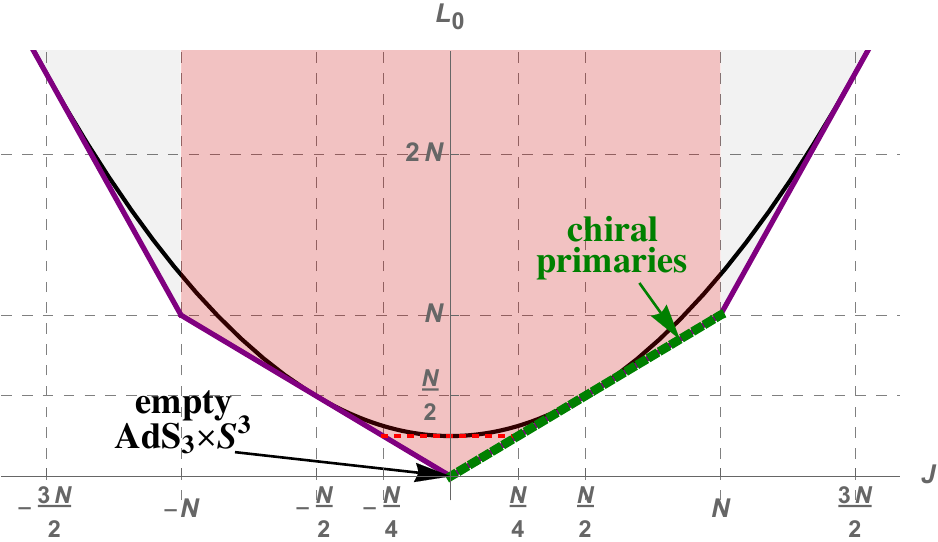}
\caption{\sl The $J$-$L_0$ plane of the 3-charge system in the NS
 sector.  The single-center BMPV black hole exists only above the black
 parabola $L_0=J^2/N+N/4$ The empty ${\rm AdS}_3\times S^3$ correspond
 to the origin $(J,L_0)=(0,0)$.  Chiral primaries live on the green
 dotted line.
  \label{fig:JNp_plane_NS}}
 \end{center}
 \end{quote}
\end{figure}

We can map these states in the R sector into the ones in the
Neveu-Schwarz (NS) sector by the spectral flow. We actually combine the
spectral flow with the $J\to -J$ symmetry and use the map
\eqref{spfl+flip} to go between the R and NS sectors.  The states in the
R sector in Figure~\ref{fig:JNp_plane_R} are mapped into the NS sector
in Figure~\ref{fig:JNp_plane_NS} where the $J$-$L_0$ diagram is shown.
The empty ${\rm AdS}_3\times S^3$ corresponds to the ground state at the
origin $(J,L_0)=(0,0)$, while 2-charge states are chiral primaries on
the line $L_0=J$ (green dashed line).

In the above, we talked about states in the range $J\in[-N/2,3N/2]$ for
the R sector and $J\in[-N,N]$ for the NS sector.  However, by spectral
flow we can get superstrata outside these ranges; see
section \ref{ss:spectral_flows} and appendix 
\ref{app:NS} for more detail.

\bigskip
The organization of the rest of the paper is as follows.
In section~\ref{sec:cft}, we review some relevant aspects of the D1-D5
CFT, in particular the structure of BPS states that can be interpreted
as the states of a gas of supergravitons in the ${\rm AdS}_3\times
S^3$ background.
In section~\ref{sec:superstrata}, we discuss superstrata solutions that
have been explicitly constructed thus far, and argue that counting
general superstrata in ${\rm AdS}_3\times S^3$ is the same as counting
the states discussed in section~\ref{sec:cft}.
In section~\ref{sec:counting}, we compute the partition function for the
superstrata in ${\rm AdS}_3\times S^3$ and estimate its entropy in the
large charge limit.  We first compute the partial contributions that
constitute the partition function and then put them together to
construct the full partition function.  See section \ref{ss:full_pf} for
the explicit formulas for the full entropy.  We find that the entropy is
parametrically smaller than the area-entropy of the D1-D5-P black hole.
In section~\ref{sec:discussion}, we speculate on what kind of gravity
configuration is relevant for the microstates of the D1-D5-P black hole.

\section{CFT side}
\label{sec:cft}

\subsection{NS sector}

Type IIB superstring on AdS$_3\times S^3\times M_4$, where $M_4=T^4$ or
K3, is dual to a $d=2,\cN=(4,4)$ CFT called the D1-D5
CFT\@.\footnote{For reviews of the D1-D5 CFT, see
e.g.~\cite{David:2002wn, Avery:2010qw}.}  The symmetry group of this
theory is $SU(1,1|2)_L\times SU(1,1|2)_R$, which is generated by the
generators $L_n,G^{\alpha A}_n,J^i_n$ and their right-moving versions
$\Lt_n,\Gt^{\dot{\alpha} A}_n,\Jt^{\,\bar{i}}_n$.  Here, $\alpha=\pm$ is
a doublet index and $i=1,2,3$ is a triplet index for $SU(2)_L\subset
SU(1,1|2)_L$, while $\dot{\alpha},\bar{i}$ are their right-moving
counterparts.  The index $A=1,2$ is the doublet index for an additional
$SU(2)_B$ symmetry group which acts as an outer automorphism on the
superalgebra.  In its moduli space, the D1-D5 CFT is believed to have an
orbifold point where the theory is described by a supersymmetric sigma
model with the target space being the symmetric orbifold ${\rm Sym}^N
M_4$.
%At the orbifold point, the
%theory is described in terms of bosons $X^i_{(r)}$ and fermions
%$\psi^i_{(r)},\psit^i_{(r)}$, where

In the NS-NS sector, the theory has one-particle chiral-primary states
which are in one-to-one correspondence with the Dolbeault cohomology of
$M_4$  \cite{Maldacena:1998bw, Vafa:1994tf}.  For $M_4=T^4$, we have 16 species of states
\begin{align}
\begin{aligned}
 &\ket{\alpha\dot\alpha}_k,&
 h&=j=\tfrac{k+\alpha}{2},& 
 \htilde&=\jt=\tfrac{k+\dot{\alpha}}{2},&&\text{bosonic},\\
 &\ket{\alpha \dot{A}}_k,&
 h&=j=\tfrac{k+\alpha}{ 2},&
 \htilde&=\jt=\tfrac{k}{ 2},&&\text{fermionic},\\
 &\ket{\dot{A}\dot\alpha}_k,&
 h&=j=\tfrac{k}{ 2},&
 \htilde&=\jt=\tfrac{k+\dot\alpha}{ 2},&&\text{fermionic},\\
 &\ket{\dot{A}\dot{B}}_k,&
 h&=j=\tfrac{k}{ 2},& 
 \htilde&=\jt=\tfrac{k}{ 2},&&\text{bosonic},
\end{aligned}
\label{T4_1-part_ch_pr}
\end{align}
where $k=1,\dots,N$.  $\dot{A},\dot{B}=1,2$ are doublet indices for an
$SU(2)_C$ that is not part of the symmetry group of the theory. $h,j$
are the values of $L_0,J_0^3$, while $\htilde,\jt$ are those of
$\Lt_0,\Jt_0^3$.  At the orbifold point, these states are twist
operators of order $k$; namely, they intertwine $k$ copies of $M_4$ (out
of $N$ copies).  We refer to these $k$ copies, thus intertwined
together, as a strand of length $k$.  Because spin is $j-\jt$, the
states $\ket{\alpha\dot\alpha},\ket{\dot{A}\dot{B}}$ are bosonic while
$\ket{\alpha\dot A},\ket{\dot{A}\dot{\alpha}}$ are fermionic.  The
$SU(2)_C$-invariant linear combination $
\tfrac{1}{\sqrt{2}}\epsilon_{\dot{A}\dot{B}}\ket{\dot{A}\dot{B}}_k$ is
denoted by $\ket{00}_k$.
For K3, there are 24 species of one-particle chiral primary states and
they are all bosonic:
\begin{align}
{\rm K3}:\qquad
\begin{aligned}
 &\ket{\alpha\dot\alpha}_k,&
 j&=\tfrac{k+\alpha}{2},& 
 \jt&=\tfrac{k+\dot{\alpha}}{2}\\
 &\ket{I}_k,&
 j&=\tfrac{k}{2},& 
 \jt&=\tfrac{k}{2},\qquad I=1,\dots,20.\\
\end{aligned}
\label{K3_1-part_ch_pr}
\end{align}
All these states \eqref{T4_1-part_ch_pr}, \eqref{K3_1-part_ch_pr}
preserve 8 supercharges, 4 from the left and another 4 from the
right.\footnote{Except for the case with $\alpha=-$ ($\dot{\alpha}=-$)
and $k=1$ for which 8 left-moving (right-moving) supercharges are
preserved.}  Conventionally, they are said to be 1/4-BPS, relative to
the amount of supersymmetry (32 supercharges) of type IIB superstring in
ten dimensions.

Among the states in \eqref{T4_1-part_ch_pr}, \eqref{K3_1-part_ch_pr},
the state $\ket{--}_1=\ket{\alpha=-,\dot\alpha=-}_1$ is special because
it has $h=j=\htilde=\jt=0$ and represents the vacuum (of a single copy
of $M_4$).  All other states can be thought of as excitations and, via
AdS/CFT, correspond to the possible excitations in linear supergravity
around empty AdS$_3\times S^3$, called supergravitons. In other words,
each of the chiral primaries \eqref{T4_1-part_ch_pr},
\eqref{K3_1-part_ch_pr} (except $\ket{--}_1$) is in one-to-one
correspondence with a particular one-particle, 1/4-BPS state of the
supergraviton propagating in the bulk AdS$_3\times S^3$ background
\cite{Deger:1998nm, Maldacena:1998bw, Larsen:1998xm, deBoer:1998kjm}.

The general chiral primary states, which are the most general 1/4-BPS states,
are obtained by multiplying together one-particle chiral primary states
as
\begin{align}
 \prod_{\psi}
 \prod_{k=1}^N
 \bigl[\ket{\psi}_k\bigr]^{N^{\psi}_k},
\label{gen_ch_pr_cft}
\end{align}
where $\ket{\psi}$ runs over different species in
\eqref{T4_1-part_ch_pr} or \eqref{K3_1-part_ch_pr}.  The general chiral
primary state is specified by the set of numbers $\{N^\psi_k\}$, which
correspond to the number of strands of species $\ket{\psi}$ and length
$k$.  The values that $N^\psi_k$ can take are $0,1,2,\dots$ if
$\ket{\psi}$ and $0,1$ if $\ket{\psi}$ is fermionic. The strand numbers
$N^\psi_k$ must satisfy the constraint that the total strand length must
be equal to $N$:
\begin{align}
 \sum_\psi \sum_k k N^\psi_{k}=N.
\label{sum_kN_k=N}
\end{align}
In the bulk, the states \eqref{gen_ch_pr_cft} correspond to
multi-particle, 1/4-BPS states of supergravitons (``supergraviton
gas'').  Namely, the states \eqref{gen_ch_pr_cft} span the Fock space of
1/4-BPS supergravitons, modulo the constraint \eqref{sum_kN_k=N}.  When
$N^{\psi}_k=\cO(N)$ (where $\ket{\psi}_k\neq \ket{--}_1$), the bulk
picture of supergravitons propagating in undeformed AdS$_3\times S^3$
is no longer valid but the geometry becomes deformed by backreaction.

The chiral primary states in \eqref{T4_1-part_ch_pr} and
\eqref{K3_1-part_ch_pr} are the highest-weight states with respect to
the rigid $SU(1,1|2)_L\times SU(1,1|2)_R$ symmetry and more general,
descendant states in the $SU(1,1|2)_L\times SU(1,1|2)_R$ multiplet can
be obtained by the action of the rigid generators
$\{L_{-1},G_{-1/2}^{-,A},J_0^-\}$ and
$\{\Lt_{-1},\Gt_{-1/2}^{-,A},\Jt_0^-\}$.  To preserve supersymmetry, we
will only consider descendants obtained by the action of the left-moving
generators $\{L_{-1},G_{-1/2}^{-,A},J_0^-\}$.  If we start with a chiral
primary with $h=j$, which we denote by $\kett{j,j}$, we generate the
following states:
\begin{align}
\begin{split}
 & \textstyle
\kett{j+n,j}\xrightarrow{J_0^-}\kett{j+n,j-1}\xrightarrow{J_0^-}\cdots\xrightarrow{J_0^-}\kett{j+n,-j}
 \\
 &G_{-1/2}^{-A}\bigg\downarrow\\
 &\textstyle
 \kett{j+\half+n,j-\half}\xrightarrow{J_0^-}\kett{j+\half+n,j-{3\over 2}}\xrightarrow{J_0^-}\cdots\xrightarrow{J_0^-}\kett{j+\half+n,-(j-\half)}
 \\
 &G_{-1/2}^{-B}\bigg\downarrow\\
 &\textstyle
 \kett{j+1+n,j-1}\xrightarrow{J_0^-}\kett{j+1+n,j-2}\xrightarrow{J_0^-}\cdots\xrightarrow{J_0^-}\kett{j+1+n,-(j-1)}
\end{split}
\label{members_of_short_multiplet}
\end{align}
Here, $\kett{h,j}$ means a state with $(L_0,J_0^3)=(h,j)$.  The states
in the second line are doubly degenerate, because we can use
$G_{-1/2}^{-A}$ with either $A=1$ or $A=2$ to descend from the first
line to the second.  The third line has no such degeneracy because we
can only descend from the first line with
$G_{-1/2}^{-,1}G_{-1/2}^{-,2}$.  More precisely, to get a genuinely new
state, we must act instead with $G_{-1/2}^{-,1}G_{-1/2}^{-,2}+{1\over
2h}L_{-1}J_0^-$ where $h$ is the value of $L_0$ for the chiral primary
\cite{Avery:2010qw,Ceplak:2018pws}.  Moreover, the number $n=0,1,\dots$
corresponds to the number of times we act on the state with $L_{-1}$.
We denote the states thus obtained building on $\ket{\psi}_k$
by\footnote{These states are not normalized.}
\begin{subequations} % 2019-04-2 11:39equations
\label{1p_1/8_sugrtn}
  \begin{align}
 \ket{\psi,k,m,n}   &=(J_0^-)^m (L_{-1})^n \ket{\psi}_k,\label{1p_1/8_sugrtn1}\\
 \ket{\psi,k,m,n,A} &=(J_0^-)^m (L_{-1})^n G_{-1/2}^{-,A} \ket{\psi}_k,\label{1p_1/8_sugrtn2}\\
 \ket{\psi,k,m,n,12}&=(J_0^-)^m (L_{-1})^n (G_{-1/2}^{-,1}G_{-1/2}^{-,2}+\tfrac{1}{2h}L_{-1}J_0^-) \ket{\psi}_k.\label{1p_1/8_sugrtn3}
 \end{align}
\end{subequations}
If the chiral primary $\ket{\psi}_k$ is bosonic (fermionic), the states
\eqref{1p_1/8_sugrtn1} and \eqref{1p_1/8_sugrtn3} are bosonic
(fermionic) while the state \eqref{1p_1/8_sugrtn2} is fermionic
(bosonic).  These states break all left-moving supersymmetry but
preserve 4 right-moving supercharges.  In the bulk, they correspond to
one-particle, 1/8-BPS supergraviton states obtained by the bulk action
of the rigid $SU(1,1|2)_L$ generators.  Actually, it is known that these
states exactly reproduce the complete spectrum of linear supergravity
around AdS$_3\times S^3$ \cite{Deger:1998nm, Maldacena:1998bw,
Larsen:1998xm, deBoer:1998kjm}.

Just as in the 1/4-BPS case, we can multiply together one-particle,
1/8-BPS states to construct a more general 1/8-BPS state:
\begin{align}
 \prod_{\psi,k,m,n,f} \bigl[\ket{\psi,k,m,n,f}\bigr]^{N^{\psi}_{k,m,n,f}},\qquad
 \sum_{\psi,k,m,n,f} k N^{\psi}_{k,m,n,f}=N,\qquad
 \label{gen_1/8_sugrtn_cft}
\end{align}
where $f={\rm null},A,12$ so it covers all the three kinds in
\eqref{1p_1/8_sugrtn}. If the state $\ket{\psi,k,m,n,f}$ is bosonic
(fermionic), $N^\psi_{k,m,n,f}=0,1,2,\dots$ ($N^\psi_{k,m,n,f}=0,1$).
The state \eqref{gen_1/8_sugrtn_cft} corresponds in the bulk to a
1/8-BPS state of the supergraviton gas.  Namely,
\eqref{gen_1/8_sugrtn_cft} spans the Fock space of 1/8-BPS
supergravitons, modulo the constraint on $N^\psi_{k,m,n,f}$.

\subsection{R sector}

By spectral flow transformation, we can map all the above statements
into the R sector.  By spectral transformation, the charges $(h,j)$ of a
state on a strand of length $k$ are transformed as follows:
\begin{align}
h'&=h+2\eta j+k\eta^2,\qquad 
j'=j+{k}\eta.\label{spfl}
\end{align}
If we take the flow parameter $\eta=-1/2$, NS states get mapped into R states.
However, to match the convention of charges to that in papers such as
\cite{Bena:2015bea, Bena:2016ypk, Bena:2017xbt}, we further flip the sign of the $SU(2)_L$ charge, as $j\to -j$. So, the
map from NS to R that we will be using is
\begin{align}
h^{\rm R}&=h^{\rm NS}-j^{\rm NS}+{k\over 4},\qquad 
j^{\rm R}={k\over 2}-j^{\rm NS}.
\label{spfl+flip}
\end{align}
The same transformation to the right-moving sector is understood.

The map \eqref{spfl+flip} transforms one-particle chiral primaries into
R ground states on a strand of length $k$.  For example,
\begin{align}
\begin{aligned}
 &\ket{--}_k^{\rm NS},&& h^{\rm NS}=j^{\rm NS}=\tfrac{k-1}{2}
 &&\to \quad
 \ket{++}_k^{\rm R},&&
 h^{\rm R}=\tfrac{k}{4},
 j^{\rm R}=\tfrac{1}{2},
 \\
 &\ket{00}_k^{\rm NS},&& h^{\rm NS}=j^{\rm NS}=\tfrac{k}{2}
 &&\to \quad
 \ket{00}_k^{\rm R},&&
 h^{\rm R}=\tfrac{k}{4},
 j^{\rm R}=0.
\end{aligned}
\end{align}
The general R ground states, which are general 1/4-BPS states, are
\begin{align}
 \prod_{\psi}
 \prod_{k=1}^N
 \left[\ket{\psi}_k^{\rm R}\right]^{N^{\psi}_k},\qquad
 \sum_\psi \sum_k k N^\psi_{k}=N.
\end{align}
where $\ket{\psi}$ runs over the species in \eqref{T4_1-part_ch_pr} or
\eqref{K3_1-part_ch_pr}, now understood as R ground states on a strand
of length $k$.  Coherent superpositions of these supergraviton states
are dual to smooth 1/4-BPS geometries called Lunin-Mathur geometries
\cite{Lunin:2001jy, Lunin:2002iz, Taylor:2005db, Kanitscheider:2007wq}, as
mentioned in the introduction.

The 1/8-BPS states in the NS sector, \eqref{gen_1/8_sugrtn_cft}, map
into the R states of the following form:
\begin{align}
 \prod_{\psi,k,m,n,f} \Bigl[\ket{\psi,k,m,n,f}^{\rm R}\Bigr]^{N^{\psi}_{k,m,n,f}},\qquad
 \sum_{\psi,k,m,n,f} k N^\psi _{k,m,n,f}=N
,\label{1/8_sugrtn_R}
\end{align}
where now the one-particle supergraviton states are given by
\begin{subequations}
\label{1p_R_states}
 \begin{align}
 \ket{\psi,k,m,n}   ^{\rm R}&=(J_{-1}^+)^m (L_{-1}-J_{-1}^3)^n \ket{\psi}_k^{\rm R},\label{1p_R_states1}\\
 \ket{\psi,k,m,n,A} ^{\rm R}&=(J_{-1}^+)^m (L_{-1}-J_{-1}^3)^n G_{-1}^{+,A} \ket{\psi}_k^{\rm R},\label{1p_R_states2}\\
 \ket{\psi,k,m,n,12}^{\rm R}&=(J_{-1}^+)^m (L_{-1}-J_{-1}^3)^n (G_{-1}^{+,1}G_{-1}^{+,2}+\tfrac{1}{2h^{\rm NS}}L_{-1}J_0^+) \ket{\psi}_k^{\rm R}.\label{1p_R_states3}
 \end{align}
\end{subequations}
The operators acting on the R ground states $\ket{\psi}_k^{\rm R}$ have
charges that have been shifted and sign-flipped due to spectral flow.
The states \eqref{1/8_sugrtn_R} are realized as superstrata in the bulk,
as we will expand in the next section.

\subsection{Higher and fractional modes}

From CFT, it is clear that the states \eqref{gen_1/8_sugrtn_cft} (or the
R version \eqref{1/8_sugrtn_R}) we constructed above are \emph{not} the
most general 1/8-BPS states, because we used only the rigid generators,
$L_{-1},G_{-1/2}^{\alpha A},J_0^i$.  We could act with higher generators
$L_{-(n+1)}, G^{\alpha A}_{-(n+1/2)}, J^i_{-n}$ with $n\ge 1$.  On a
stand of length $k>1$, we could also act with fractional generators
$L_{-{n\over k}}, G^{\alpha A}_{-{n+1/2\over k}}, J^i_{-{n\over k}}$.
However, these are not included in the states~\eqref{gen_1/8_sugrtn_cft}
(or \eqref{1/8_sugrtn_R}, with the mode numbers of the generators
appropriately shifted).  Actually, it is known that, once we turn on
perturbation and leave the orbifold point of the D1-D5 CFT
\cite{Gava:2002xb}, many of such higher/fractional states lift (see
\cite{Guo:2019pzk} for a more recent discussion) and disappear from the
BPS spectrum.  Nevertheless, there are ones that do not lift and are
important to account for the entropy of the 3-charge black hole.

Actually, we can ask if the supergraviton states
\eqref{gen_1/8_sugrtn_cft} (or \eqref{1/8_sugrtn_R}) also disappear when
we leave the orbifold point of the D1-D5 CFT\@.
For $M={\rm K3}$, we can see that these supergraviton states do not
lift because they make non-vanishing contribution to the supersymmetric
index (elliptic genus) \cite{deBoer:1998us}.  This in particular means
that, at any point in the moduli space of K3 surfaces,\footnote{The
moduli space of K3 surfaces is part of the moduli space of the D1-D5
CFT, which also includes moduli corresponding to the NS-NS and RR fields
through the K3.} there are 1/8-BPS supergravitons in linear supergravity
around ${\rm AdS}_3\times S^3\times {\rm K3}$.
For $M=T^4$, on the other hand, we cannot use an argument based on the
supersymmetry index (the modified elliptic genus) because supergravitons
do not contribute to it (at least for $L_0^{\rm NS}< N/4$)
\cite{Maldacena:1999bp}.
However, recall that, at the orbifold points in the moduli space of K3
surfaces, K3 is realized as $T^4/\bbZ_l$ ($l=2,3,4,6$)
\cite{Walton:1987bu}.\footnote{These orbifold points in the moduli space
of K3 surfaces are not to be confused with the orbifold points in the
moduli space of the D1-D5 CFT where the CFT target space is a symmetric
orbifold of the K3 surface.}  As we just mentioned, the BPS spectrum of
linear supergravity for this ${\rm K3}=T^4/\bbZ_l$ is non-empty.  This
means that the BPS spectrum of linear supergravity for the parent $T^4$
is also non-empty and presumably completely unlifted.\footnote{Some
modes (the ones associated with the fermionic chiral primaries) in the
$T^4$ spectrum gets projected out by the $\bbZ_l$ orbifold action, while
the K3 spectrum contains extra modes that come from the 16 collapsed
2-cycles at the fixed points of the orbifold.  The overlap in the
spectra (the modes associated with the bosonic chiral primaries of
$T^4$) do not lift.  The fact that the modified elliptic genus for $T^4$
vanishes most likely means that the modes associated with the fermionic
chiral primaries do not lift either, responsible for the cancellation.}
Namely, even for $M_4=T^4$, there exist BPS supergravitons and thus
superstrata that are counted in this paper, even away from the orbifold
point, for the values of the $T^4$ moduli for which $T^4$ can be
orbifolded to give K3 \cite{Walton:1987bu}.

\section{Superstrata}
\label{sec:superstrata}

Superstrata are regular horizonless 1/8-BPS solutions of supergravity
that describe microstates of the D1-D5-P black hole.  They are dual to
coherent superpositions of the 1/8-BPS states~\eqref{1/8_sugrtn_R} of
the supergraviton gas and fully take into account the backreaction for
$N^\psi_{k,m,n,f}=\cO(N)$.

More precisely,  the superstrata that have
been explicitly constructed thus far
are believed to involve macroscopic excitation of the following species of
one-particle supergravitons:\footnote{Actually, there are
superstrata that do not correspond to supergraviton states
\eqref{1/8_sugrtn_R} obtained by exciting supergravitons around
AdS$_3\times S^3$.  In \cite{Bena:2016agb}, superstrata solutions were
written down that are obtained by exciting supergravitons around an
orbifold background $({\rm AdS}_3\times S^3)/\bbZ_k$.  The dual CFT
states involve certain fractional generators and cannot be written in
the form~\eqref{1/8_sugrtn_R}.  }
\begin{itemize}
 \item $\ket{00,k,m,n}^{\rm R}$: the ``original'' superstratum
       \cite{Bena:2015bea, Bena:2016ypk, Bena:2017xbt}.
 \item $\ket{++,k,m,n}^{\rm R},\ket{--,k,m,n}^{\rm R}$: ``style 1''
       \cite{Bena:2016agb}.
 \item $\ket{\dot{A}\dot{B},k,m,n}^{\rm R}$: superstratum with internal
       excitations\cite{Bakhshaei:2018vux}.
 \item $\ket{00,k,m,n,12}^{\rm R}$: supercharged superstratum
       \cite{Ceplak:2018pws}.
 \item $\ket{00,k,m,n}^{\rm R}, \ket{00,k,m,n,12}^{\rm R}$: ``hybrid''
       superstratum \cite{Heidmann:2019zws}.
\end{itemize}
In addition, these all involve $\ket{++,1,0,0}^{\rm R}$ which
corresponds to the vacuum in the NS sector.

These superstrata solutions were constructed relying on a linear
structure that the BPS equations of 6D supergravity possess
\cite{Bena:2011dd}.  In particular, these solutions have a flat
four-dimensional base on which the BPS equations are defined.  If one
considers more general species, such as $\ket{+-}$, the base gets
deformed and it becomes more difficult to construct backreacted
supergravity solutions by making use of the linear structure.
However, this is only a technical issue, and it is physically natural to
expect that, for every species $\ket{\psi,k,m,n,f}^{\rm R}$, there
exists a smooth backreacted solution (superstratum) that corresponds to
a macroscopic, coherent excitation of the corresponding supergraviton.
Below, we assume that we can identify the states~\eqref{1/8_sugrtn_R}
with such general superstrata, and that geometric quantization of
the latter will reproduce the Hilbert space of
states~\eqref{1/8_sugrtn_R}.  Under the assumption, we now count
superstrata by counting the states~\eqref{1/8_sugrtn_R}.

Note that the ``stringy exclusion principle'' constraint
\cite{Maldacena:1998bw} such as the second equation of
\eqref{1/8_sugrtn_R} has been observed to be automatically imposed in
fully backreacted geometries, such as superstrata \cite{Bena:2015bea},
as regularity constraint \cite{deBoer:2009un}.  Therefore, we assume
that counting the states~\eqref{1/8_sugrtn_R} including
the constraint is the same as counting all 
regular
superstrata.

\section{Counting}
\label{sec:counting}

In this section, we carry out the counting of the number of
states~\eqref{1/8_sugrtn_R}, which we assume to be the same as counting
the dual superstrata solutions.
More precisely, we will compute the 1/8-BPS grand-canonical partition function
\begin{align}
 Z(p,q,y)=\sum_{N=0}^\infty p^N Z_N(q,y),\qquad
 Z_N(q,y)=\Tr_{\text{R,1/8-BPS}}[q^{h-N/4}  y^j]
\label{def_Z_R}
\end{align}
where $Z_0(q,y)\equiv 1$ and the trace is over the Fock space states in
the R sector, \eqref{1/8_sugrtn_R}, for given~$N$\@.  We do not have to consider the
constraint on the total strand length in the grand-canonical partition
function since it is taken care of by the strand-length counting
parameter~$p$.
Note that, due to the shift by $-N/4$ in the definition \eqref{def_Z_R},
the power of $q$ records not the value of~$L_0$ but 
\begin{align}
 N_p=L_0-{N\over 4},
\end{align}
which vanishes for R ground states. 

\subsection{The partition function}

To compute $Z$, it is perhaps easiest to go back to the NS sector and
start with the contribution of the states in
\eqref{members_of_short_multiplet} to an NS sector grand-canonical
partition function.  If the chiral primary~$\ket{\psi}_k^{\rm NS}$ is a
bosonic state, the contribution to the partition function from it and
from its descendants is
\begin{align}
z^{\rm NS, bos}_{j,\tilde{j},k}
 &=
 \prod\limits_{n=0}^\infty
 {
 \prod_{i=-(j-\half)}^{j-\half} (1+p^k q^{j+\half+n} y^i)^2
 \over 
 \prod_{i=-j}^j (1-p^k q^{j+n} y^i)
 \prod_{i=-(j-1)}^{j-1} (1-p^k q^{j+1+n} y^i)
 }.
\label{z_NS_bos}
\end{align}
The two factors in the denominator come from the first and third lines
of \eqref{members_of_short_multiplet}, while the numerator comes from
the second line of \eqref{members_of_short_multiplet}.  The product over
$i$ comes from the action of $J_0^-$ and the product of $n$ from the
action of $L_{-1}$.  This \eqref{z_NS_bos} represents the contribution
of multi-particle states of supergravitons, built on $\ket{\psi}_k$ with
the particular string length, $k$.
By using the standard technique of taking the logarithm and using the
formula $-\log(1-x)=\sum_{r=1}^\infty x^r/r$, we can carry out the
summation over $i$ and $n$, obtaining
\begin{align}
\log z^{\rm NS, bos}_{j,\tilde{j},k}
&=
  \sum_{r=1}^\infty{p^{rk} q^{rj}\over r(1-q^r)(1-y^r)}\Bigl[
 y^{-rj}(1-2(-1)^rq^{r/2}y^{r/2}+q^r y^r)\notag\\[-2ex]
 &\hspace{30ex}
 -y^{rj}(y^r-2(-1)^rq^{r/2}y^{r/2}+q^r)
 \Bigr].
\label{log_z_NS_bos}
\end{align}
Note that this does not depend on $\tilde j$.  This expression is valid
for states built on bosonic chiral primaries in \eqref{T4_1-part_ch_pr}
and \eqref{K3_1-part_ch_pr}, namely $\ket{\alpha\dot\alpha}_k$,
$\ket{\dot{A}\dot{B}}_k$, and $\ket{I}_k$, which have
\begin{align}
 \textstyle
 j^{\rm NS}={k+a\over 2},  \qquad a=\pm, 0.
\end{align}
Summing over all possible strand length, $k=1,2,\dots$, we obtain the
contribution from the states of species
$\ket{\psi}=\ket{\alpha\dot\alpha}, \ket{\dot{A}\dot{B}}, \ket{I}$:
\begin{align}
\log z^{\rm NS, bos}_{\ket{a}}
&\equiv
\sum_{k=1}^\infty \log z^{\rm NS, bos}_{{k+a\over 2},\tilde{j},k}
\notag\\
&=
\sum_{r=1}\limits^\infty{p^{r}q^{{1+a\over 2}r} \over r(1-q^r)(1-y^r)}\biggl[
 {y^{-{1+a\over 2}r}(1-2(-1)^r q^{r\over 2}y^{r\over 2}+q^r y^r)
 \over 1-(p q^{1\over 2}y^{-{1\over 2}})^r}\notag\\
 &\hspace{30ex}
 -{y^{{1+a\over 2}r}(y^r-2(-1)^r q^{r\over 2}y^{r\over 2}+q^r)
   \over 1-(p q^{1\over 2}y^{1\over 2})^r}
 \biggr].\label{z_NS_bos_a}
\end{align}
Here, we used the notation $z^{\rm NS, bos}_{\ket{a}}$ because this
depends only on the value of $a$ of the chiral primary.  We will often
suppress the right-moving part of the states, which is irrelevant, and
write e.g.~$\ket{+\pm}$ as $\ket{+}$ and $\ket{00}$ as $\ket{0}$.

Let us translate this into the R sector.  By inspecting the
transformation \eqref{spfl+flip}, one sees that the NS partition
function can be converted into the R partition function by the
replacement
\begin{align}
 p\to p y^{{1\over 2}},\qquad
 q\to q,\qquad
 y\to q^{-1}y^{-1}.
\end{align}
Here, we have also taken into account the fact that we have defined the
R partition function~\eqref{def_Z_R} with the shift $h\to h-N/4$.  By applying this to \eqref{z_NS_bos_a}, we find that 
the contribution to $Z(p,q,y)$ from states built on the bosonic R ground
state $\ket{\psi}^{\rm R}$ with $j={a\over 2}$ is
\begin{align}
  \log z^{\rm bos}_{\ket{a}}
 &=
 \sum\limits_{r=1}^\infty{p^{r} y^{a r\over 2} \over r(1-q^r)(1-q^{r}y^{r})}\left[
 {1-2(-1)^r q^r y^{r\over 2}+q^{2r}y^r   \over 1-p^r}
 -
 { q^{(2-a)r} y^{(1-a)r} (1-2(-1)^r y^{r\over 2}+y^{r})
 \over 1-(p q y)^{r}}
 \right].
\label{z_R_bos_a}
\end{align}
where we omitted the superscript ``R'' for the R sector, because we
will only be discussing R sector quantities henceforth.

Similarly, if the R ground state $\ket{\psi}^{\rm R}$ is fermionic
and has $j={a\over 2}$, the
contribution from states built on it is
\begin{align}
 \log z^{\rm fer}_{\ket{a}}
 &=
 -\sum\limits_{r=1}^\infty{p^{r} y^{a r\over 2} \over r(1-q^r)(1-q^{r}y^{r})}\biggl[
 {(-1)^r-2 q^r y^{r\over 2}+(-1)^rq^{2r}y^r   \over 1-p^r}
 \notag\\
 &\hspace{30ex}
 -
 {q^{(2-a)r} y^{(1-a)r} ((-1)^r-2 y^{r\over 2}+(-1)^ry^{r})
 \over 1-(p q y)^{r}}
 \biggr].
\label{z_R_fer_a}
\end{align}

Summing \eqref{z_R_bos_a} and \eqref{z_R_fer_a} over all the species in
\eqref{T4_1-part_ch_pr} and \eqref{K3_1-part_ch_pr},
we finally obtain the grand-canonical partition function for
$T^4$ and K3  as follows:
\begin{subequations}
 \label{ZT4,ZK3_as_sum}
\begin{align}
 Z_{T^4}(p,q,y) 
 &=
  2\log z^{\rm bos}_{\ket{+}}
 +4\log z^{\rm bos}_{\ket{0}}
 +2\log z^{\rm bos}_{\ket{-}}
 +2\log z^{\rm fer}_{\ket{+}}
 +4\log z^{\rm fer}_{\ket{0}}
 +2\log z^{\rm fer}_{\ket{-}},
 \\[2ex]
 Z_{\rm K3}(p,q,y) 
 &=2\log z^{\rm bos}_{\ket{+}}
 +20\log z^{\rm bos}_{\ket{0}}
 + 2\log z^{\rm bos}_{\ket{-}}.
 \end{align}
\end{subequations}
Our goal is to estimate these partition functions to estimate the number
of states.

\subsection{Thermodynamic approximation}

Let us expand the partition function
$Z(p,q,y)$ as
\begin{align}
 Z(p,q,y)=\sum_{N,N_p,J} c(N,N_p,J)\,p^N q^{N_p} y^J.
\end{align}
What we would like to know is the behavior of the coefficient
$c(N,N_p,J)$ in the large charge limit,
\begin{align}
 N\gg 1, \qquad N_p\gg 1, \qquad
 N_p=\cO(N),\qquad |J|=\cO(N).
\label{large_chg_lim}
\end{align}
Note that $N$ and $N_p$ are large and positive, while $|J|$ can be as
large as $N\sim N_p$.  So, $J$ can be either positive, negative, or zero.
We introduce the chemical potentials $\alpha,\beta,\gamma$ by
\begin{align}
 p=e^{-\alpha},\qquad q=e^{-\beta},\qquad y=e^{-\gamma},
\label{pqr_alpha,beta,gamma}
\end{align}
where $\alpha,\beta>0$ while $\gamma$ can be positive, negative, or
zero.  In the limit \eqref{large_chg_lim}, we have
\begin{align}
 \alpha,\beta, |\gamma|=\cO( \epsilon),\qquad \epsilon \to 0,
 \label{small_alpha,beta,gamma}
\end{align}
where $\epsilon>0$ indicates how fast $\alpha,\beta,\gamma$
are going to zero.  In the large charge limit, we can use thermodynamic
approximation, within which we have
\begin{align}
 N=-\partial_\alpha  \log Z,\qquad
 N_p=-\partial_\beta  \log Z,\qquad
 J=-\partial_\gamma  \log Z\label{N,Np,J_alpha,beta,gamma}
\end{align}
and entropy is given by
\begin{align}
 S(N,N_p,J)=\log c(N,N_p,J)
 =\log Z+\alpha N+\beta N_p + \gamma J.\label{S_N,Np,J_alpha,beta,gamma}
\end{align}

\subsection{Warm-up: a 2-charge system with angular momentum}
\label{ss:2-chg_counting}

Before estimating the 3-charge partition function of supergravitons, it
is instructive to look at 2-charge partition function to gain an idea.

Consider a 2-charge ensemble made of states of the form
\begin{align}
 \prod_{k=1}^{N} \bigl[\ket{+}_k\bigr]^{N_k},\qquad \sum_{k} k N_k=N.
\end{align}
Namely, we have only one species of strands, carrying angular momentum
$j=\half$, with no excitation on top.
The partition function for this system is
\begin{align}
 Z_{\text{2-chg}}(p,y)=\prod_{k=1}^\infty {1\over 1-p^ky^{1/2}}.\label{jigf31May19}
\end{align}
If we also included $\ket{-}$ strands, this could be related to theta
functions and could be estimated using modular properties of theta
functions \cite{Russo:1994ev}, but we will not do so but take a
thermodynamic approach.

As before, \eqref{jigf31May19} can be written as
\begin{align}
 \log Z_{\text{2-chg}}(p,y)
 =\sum_{r=1}^\infty {(p y^{1/2})^r \over r(1-p^r)}.\label{jjjk31May19}
\end{align}
In the large $N,J$ limit where $\alpha,|\gamma|\ll 1$, one may wonder if we can simply plug in
\eqref{pqr_alpha,beta,gamma} to obtain
\begin{align}
 \log Z_{\text{2-chg}}(p,y)
 \stackrel{?}{\approx}
 \sum_{r=1}^\infty {1\over r^2\alpha}
 = {\pi^2\over 6\alpha}.
\end{align}
However, this is incorrect as the formula
\eqref{N,Np,J_alpha,beta,gamma} would always give $J=0$.

What went wrong is physically clear: if $J>0$, the strand $\ket{+}_1$
condenses, because that is the most economical way to carry angular
momentum.  This is manifested in \eqref{jjjk31May19} in the fact that
the sum diverges if $py^{1/2}=e^{-(\alpha+\gamma/2)}\to 1$, which
happens if $\gamma\to -2\alpha+0$.  If we extract the dangerous
contribution from the partition function \eqref{jigf31May19} and rewrite
the rest as an $r$-sum, we find
\begin{align}
 \log Z_{\text{2-chg}}(p,y)
 =-\log(1-py^{1/2}) + \sum_{r=1}^\infty  {(p^2 y^{1/2})^r \over r(1-p^r)}
 \approx -\log\left(\alpha+{\gamma\over 2}\right) +{\pi^2\over 6\alpha}.
\end{align}
Note that $p^2 y^{1/2}=e^{-2\alpha-\gamma/2}\approx e^{-2\alpha}$ is not
dangerous even if $\gamma\to -2\alpha$.  Then, by a straightforward
application of the thermodynamic formulas, we find
\begin{align}
 \alpha={\pi\over \sqrt{6(N-J)}},\qquad
 \alpha+{\gamma\over 2}={1\over J },\qquad
\end{align}
Therefore, $\alpha,\gamma=\cO(\epsilon)$, while
$\alpha+\gamma/2=\cO(\epsilon^2)$. The entropy is
\begin{align}
 S_{\text{2-chg}}=2\pi\sqrt{N-J\over 6},\qquad N\ge J> 0,\label{S_2-chg}
\end{align}
where we dropped subleading terms (such as $\log
J$).  This is a well-known 2-charge entropy formula \cite{Iizuka:2005uv}.

Some comments are in order.
(i) In this example, because all strands have $J>0$, there is no state
with $J=0$ and therefore it must be that $S_{\text{2-chg}}(N,J)\to 0$ as
$J\to 0$. However, this is invisible in~\eqref{S_2-chg}, which is
valid only in the large $N\sim J$ limit.
(ii) We said that a condensate of $\ket{+}_1$ absorbs $J$.  Strictly
speaking, we do not have $[\ket{+}_1]^{2J}_{}$ but
$[\ket{+}_1]^{2(J-x)}_{}$ with $1\ll x\ll J$. The
remaining strand-length budget of order $N-2(J-x)\approx N-2J$ is occupied
by $2x$ $\ket{+}_k$ strands of average length $k\sim (N-2J)/2x$ and is
responsible for the entropy.  The spin carried by these strands ($\sim
x$) is negligible compared to that of the condensate ($\sim J$).  One
can show that $x=\cO(\sqrt{N-2J}\log(N-2J))$.
(iii) More generally, if we have $c_\pm$ species of $\ket{\pm}$ carrying $j=\pm\half$
and $c_0$ species of $\ket{0}$ with $j=0$,  entropy becomes
$S_{\text{2-chg}}=2\pi\sqrt{c(N-|J|)/ 6}$, $c\equiv c_+ + c_- + c_0$.
Depending on $J\gtrless 0$, the strand $\ket{\pm}_1$ condenses.

\subsection{Estimating building blocks}

Now let us come back to the estimation of the 3-charge partition function of
super\-strata/super\-gravitons.  Before estimating the entropy for the full
partition function \eqref{ZT4,ZK3_as_sum}, it is useful to first
estimate the individual contributions, such as $z^{\rm bos}_{\ket{0}},
z^{\rm bos}_{\ket{+}}, z^{\rm bos}_{\ket{-}}$.

\subsubsection{Case 1: $z^{\rm bos}_{\ket{0}}$}
\label{sss:z_bos_0}

Let us start with $z^{\rm bos}_{\ket{0}}$.    Namely, we expand
$z^{\rm bos}_{\ket{0}}$ as
\begin{align}
 z^{\rm bos}_{\ket{0}}
 =\sum_{N,N_p,J}c^{\rm bos}_{\ket{0}}(N,N_p,J)\,p^N q^{N_p} y^J
\end{align}
and want to estimate the behavior of $S^{\rm bos}_{\ket{0}}=\log c^{\rm
 bos}_{\ket{0}}(N,N_p,J)$ in the large charge limit.

If we set $a=0$ in
\eqref{z_R_bos_a}, plug in \eqref{pqr_alpha,beta,gamma}, and 
expand the summand according to \eqref{small_alpha,beta,gamma}, we obtain
\begin{align}
 \log z^{\rm bos}_{\ket{0}}
=
 \sum_{r=1}^\infty \left[
 {2(1-(-1)^r) \over r^4 \alpha \beta (\alpha+\beta+\gamma)}
 +\cO(\epsilon^{-2})\right].
\end{align}
Ignoring the $\cO(\epsilon^{-2})$ terms and carrying out the summation over $r$,
we obtain
\begin{align}
 \log z^{\rm bos}_{\ket{0}}
 ={C\over \alpha' \beta' \gamma'},
\qquad
 C\equiv{\pi^4\over 24},\label{zRbos0_ito_a'b'g'}
\end{align}
where we have defined
\begin{align}
 \alpha'\equiv\alpha,\qquad
 \beta'\equiv\beta,\qquad
 \gamma'\equiv\alpha+\beta+\gamma.
\end{align}
If we introduce $N',N_p',J'$ conjugate to $\alpha',\beta',\gamma'$ by
\begin{align}
 N'=N-J,\qquad
 N_p'=N_p-J,\qquad
 J'=J\label{def_N'Np'J'}
\end{align}
then the thermodynamic relations \eqref{N,Np,J_alpha,beta,gamma} and
\eqref{S_N,Np,J_alpha,beta,gamma} become
\begin{align}
 N'=-{\p_{\alpha'}\log Z},\quad
 N_p'=-{\p_{\beta'}\log Z},\quad
 J'=-{\p_{\gamma'}\log Z}
\label{N'Np'J'_a'b'g'}
\end{align}
and
\begin{align}
 S=\log Z+\alpha'N'+\beta'N_p'+\gamma'J'.
\end{align}

\begin{figure}[htb]
 \begin{quote}
 \begin{center}
     \includegraphics[height=5cm]{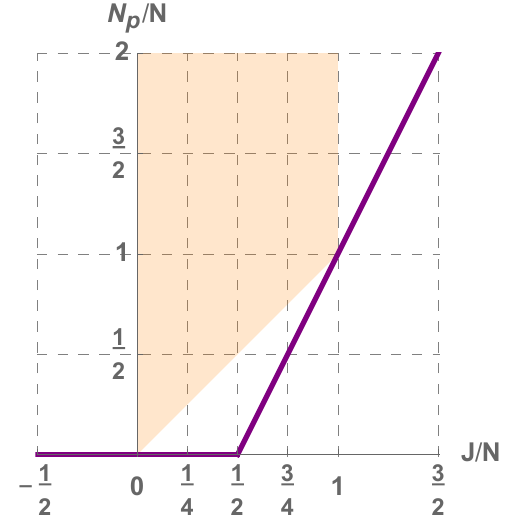}
\caption{\sl The region in which states counted by $z^{\rm
  bos}_{\ket{0}}$ exist.   The purple lines
  represent the unitarity bound below which no states exist.
  \label{fig:regions0}}
 \end{center}
 \end{quote}
\end{figure}

In the present case, using \eqref{zRbos0_ito_a'b'g'} and
\eqref{N'Np'J'_a'b'g'}, we find
\begin{align}
 N'={C\over \alpha'^2\beta'\gamma'},\qquad
 N_p'={C\over \alpha'\beta'^2\gamma'},\qquad
 J'={C\over \alpha'\beta'\gamma'^2}.\label{nkcy6May19}
\end{align}
and the entropy is
\begin{align}
S^{\rm bos}_{\ket{0}}
 = {4C \over \alpha'\beta'\gamma'}.
\end{align}
Because $\alpha',\beta'>0$, in order for the entropy $S^{\rm
bos}_{\ket{0}}$ to be positive, we need $\gamma'>0$ and therefore
$N',N_p'\ge 0$. It is also clear that $J'\ge 0$.  In terms of $N,N_p,J$, we
 need
\begin{align}
 N\ge J, \qquad N_p\ge J,\qquad J\ge 0.
\end{align}
See Fig.~\ref{fig:regions0} for the region where $c^{\rm bos}_{\ket{0}}(N,N_p,J)\neq 0$.

The relations \eqref{nkcy6May19} can  be solved for $N',N_p',J'$ as
\begin{align}
 \alpha'= \left({C N_p' J'\over N'^3} \right)^{1/4},\qquad
 \beta' = \left({C N' J'\over  N_p'{}^3} \right)^{1/4},\qquad
 \gamma'= \left({C N' N_p'\over J'^3} \right)^{1/4},\label{a'b'g'_ito_N'Np'J'_0}
\end{align}
which gives the entropy of $z^{\rm bos}_{\ket{0}}$ in terms of charges
to be
\begin{align}
S^{\rm bos}_{\ket{0}}
% = {4C \over \alpha'\beta'\gamma'}
 = 4(C N' N_p' J')^{1/4}
 =4\pi \left[{1\over 24}J (N - J) (N_p-J)\right]^{1/4}.\label{SRbos_0}
\end{align}

\subsubsection{Case 2: $z^{\rm bos}_{\ket{+}}$}
\label{sss:z_bos_+}

Let us turn to $z^{\rm bos}_{\ket{+}}$.  Just as in the 2-charge case
studied in section \ref{ss:2-chg_counting}, in an ensemble of strands
with angular momentum, condensation of the shortest-length strand is
expected when $J$ is large (see \cite{Bena:2011zw, Iizuka:2005uv}).
We will indeed see such condensation in this case.

By setting $a=+1$ in \eqref{z_R_bos_a},
we find
\begin{align}
  \log z^{\rm bos}_{\ket{+}}
 &=
 \sum\limits_{r=1}^\infty{p^{r} y^{r\over 2} \over r(1-q^r)(1-q^{r}y^{r})}\left[
 {1-2(-1)^r q^r y^{r\over 2}+q^{2r}y^r   \over 1-p^r}
 -
 { q^{r} (1-2(-1)^r y^{r\over 2}+y^{r})
 \over 1-(p q y)^{r}}
 \right].
 \label{z_R_bos_a=+}
\end{align}
If we naively do small-$\epsilon$ expansion of the summand, as we did
for $z^{\rm bos}_{\ket{0}}$, then we get the same expression
\eqref{zRbos0_ito_a'b'g'} and the same entropy \eqref{SRbos_0}.
However, such expansion is not always valid.  In the numerator
of~\eqref{z_R_bos_a=+}, we have $p^r
y^{r\over 2}=(e^{-\alpha-\gamma/2})^r=(e^{-{1\over
2}(\alpha'-\beta'+\gamma')})^r$.  So, for the series to converge, we
need
\begin{align}
 \alpha'-\beta'+\gamma'
 ~\propto~ N_p'J'-N'J'+N'N_p'
 ~=~ J^2-2JN+NN_p>0,
\label{irbd5Apr19}
\end{align}
where in the ``$\propto$'' we used \eqref{a'b'g'_ito_N'Np'J'_0}.
However, this inequality is not always satisfied.\footnote{On the other
hand, if it is satisfied, other quantities in the numerator, such as
$p^r y^{r/2} \cdot q^{2 r} y^r$ are all strictly smaller than one and
the series converges.}$^{,}$\footnote{More precisely, the left-hand side
can go to zero faster than $\epsilon$, just as for the 2-charge case
studied in section \ref{ss:2-chg_counting}.}
When the left-hand side of \eqref{irbd5Apr19} becomes very small, the
expansion of the summand that led to~\eqref{zRbos0_ito_a'b'g'} is no
longer valid.  This is because the state counted by $p^r y^{r/2}=(p
y^{1/2})^r$, namely, $\ket{+}_1$, condenses.

Let us focus on the first term in \eqref{z_R_bos_a=+} and extract terms
that become large when the left-hand side of \eqref{irbd5Apr19} becomes
very small.  Using $p^r=p^r(1-p^r)+p^{2r}$, we can rewrite it as
\begin{align}
  \sum_{r=1}^\infty
 \frac{p^r y^{r\over 2} }{r (1-p^r)   (1-q^r) (1-q^r y^r)}
 =
  \sum_{r=1}^\infty
 \frac{p^r y^{r\over 2} }{r   (1-q^r) (1-q^r y^r)}
 +
  \sum_{r=1}^\infty
 \frac{p^{2r} y^{r\over 2} }{r (1-p^r)   (1-q^r) (1-q^r y^r)}.
\label{glei7Apr19}
\end{align}
The second term on the right is safe, as long as $p<1$.  On the other
hand, the first term can be rewritten as
\begin{align}
 \sum_{r=1}^\infty
 \frac{p^r y^{r\over 2} }{r 
   (1-q^r) (1-q^r y^r)}
 =
 \sum_{r=1}^\infty
 \sum_{l=0}^\infty \sum_{n=0}^\infty
 \frac{p^r y^{r\over 2} q^{n r} (q y)^{l r}}{r}.
\end{align}
The $l=n=0$ terms give:
\begin{align}
 \sum_{r=1}^\infty
 \frac{p^r y^{r\over 2}}{r}
&=
 -\log(1-py^{1/2})
\approx
 -\log{\alpha'-\beta'+\gamma\over 2},
\end{align}
while other ($(l,n)\neq (0,0)$) terms give $\cO(\epsilon^{-2})$
contributions which are subleading compared to \eqref{zRbos0_ito_a'b'g'}.
The remaining terms in~\eqref{z_R_bos_a=+} and the second term in~\eqref{glei7Apr19} give \eqref{zRbos0_ito_a'b'g'}, just as for the
case with $z^{\rm bos}_{\ket{0}}$.  So, the partition function is given by
\begin{align}
 \log z^{\rm bos}_{\ket{+}}
 \approx 
 -\log{\alpha'-\beta'+\gamma'\over 2}+{C\over \alpha'\beta'\gamma'}.\label{hpcy7May19}
\end{align}

Using thermodynamic relations \eqref{N'Np'J'_a'b'g'}, we obtain
\begin{align}
  N'
 =\Delta^{(1)}+{C\over \alpha'^2\beta'\gamma'},\qquad
 N_p'
 =-\Delta^{(1)}+{C\over \alpha'\beta'^2\gamma'},\qquad
 J'
 =\Delta^{(1)}+{C\over \alpha'\beta'\gamma'^2},
\label{goyc7Apr19}
\end{align}
where
\begin{align}
 \Delta^{(1)}\equiv{1\over \alpha'-\beta'+\gamma'}>0.\label{Delta1}
\end{align}
The two terms on the right-hand side of \eqref{goyc7Apr19} are of the
same order if
\begin{align}
 \alpha',\beta',\gamma'\sim \epsilon
 \qquad \text{but} \qquad
 \alpha'-\beta'+\gamma'\sim \epsilon^4.
\end{align}
The entropy is computed to be
\begin{align}
 S^{\rm bos}_{\ket{+}}=\log Z+\alpha'N'+\beta'N_p'+\gamma'J'
 \approx {4C\over \alpha'\beta'\gamma'},
\end{align}
where we have dropped subleading terms, including the $\cO(\log \epsilon)$
term that comes from the first term in \eqref{hpcy7May19}.

Looking at the definition of $N',N_p',J'$ in \eqref{def_N'Np'J'}, we see
that the existence of the $\Delta^{(1)}$ terms in~\eqref{goyc7Apr19}
change $N$ by $2\Delta^{(1)}$ and $J$ by $\Delta^{(1)}$.  This means
that we have the following condensate:
\begin{align}
 \bigl[\ket{+}_1\bigr]^{2\Delta^{(1)}}.
\end{align}

\begin{figure}[htb]
 \begin{quote}
 \begin{center}
     \includegraphics[height=5cm]{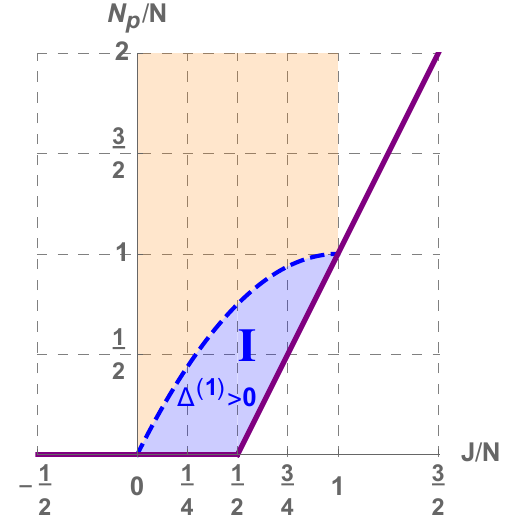}
\caption{\sl The regions in which states counted by $z^{\rm
  bos}_{\ket{+}}$ exist (orange and blue regions).  In the blue region,
  which is given by \eqref{regionI}, the strand $\ket{+}_1$ condenses
  and $\Delta^{(1)}>0$.  On the dashed line, $\Delta^{(1)}=0$.
  \label{fig:regions+}}
 \end{center}
 \end{quote}
\end{figure}

In order to express the entropy in terms of charges, let us define
\begin{align}
\begin{split}
 N^{(1)}&\equiv N'-\Delta^{(1)}
 %=N-J-\Delta^{(1)}
 ,\qquad
 N_p^{(1)}\equiv N_p'+\Delta^{(1)}
 %=N_p-J+\Delta^{(1)}
 ,\qquad
 J^{(1)}\equiv J'-\Delta^{(1)}%=J-\Delta^{(1)}
 .
\end{split}
\label{hdcw10Apr19}
\end{align}
Then \eqref{goyc7Apr19} become
\begin{align}
 N^{(1)}={C\over \alpha'^2\beta'\gamma'},\qquad
 N_p^{(1)}={C\over \alpha'\beta'^2\gamma'},\qquad
 J^{(1)}={C\over \alpha'\beta'\gamma'^2},
\end{align}
which have the same form as
\eqref{nkcy6May19} and we immediately obtain
\begin{align}
 \alpha'=\biggl({C N_p^{(1)}J^{(1)}\over (N  ^{(1)})^3}\biggr)^{1/4},\qquad
 \beta' =\biggl({C N^{(1)}J^{(1)}  \over (N_p^{(1)})^3}\biggr)^{1/4},\qquad
 \gamma'=\biggl({C N^{(1)}N_p^{(1)}\over (J  ^{(1)})^3}\biggr)^{1/4}
\label{hdbs10Apr19}
\end{align}
and
\begin{align}
 S^{\rm bos}_{\ket{+}}=4\bigl(C N^{(1)} N_p^{(1)} J^{(1)}\bigr)^{1/4}.
\label{hnes10Apr19}
\end{align}
On the other hand, the condition $\alpha'-\beta'+\gamma'\approx 0$ means
that
\begin{align}
 N_p^{(1)} J^{(1)} - N^{(1)} J^{(1)} + N^{(1)} N_p^{(1)} = 0,
\end{align}
which we can solve to find $\Delta^{(1)}$ in terms of $N,N_p,J$.
The solution is
\begin{align}
 \Delta^{(1)}=\frac{1}{3} 
 \left(J-N_p+N-\sqrt{4 J^2-2 J (N_p+2N)+N_p^2+N N_p+N^2}\right).
\label{jrrv10Apr19}
\end{align}
With this sign choice in front of the square root, the region in which
$\Delta^{(1)}\ge 0$ is
\begin{align}
 N_p\le J \left(2 - {J\over N}\right),\qquad
 N_p\ge 0,\qquad N_p\ge 2J-N\qquad
 \text{(region I)}.
 \label{regionI}
\end{align}
We also included inequalities (the second and third ones) which come
from the condition $N^{(1)},N_p^{(1)},J^{(1)}\ge 0$ required for the
entropy to be positive, as in the case for $z^{\rm bos}_{\ket{0}}$.
This region~\eqref{regionI} is in the complement of the region given
by~\eqref{irbd5Apr19}.  The condensate $\Delta^{(1)}$ vanishes when the
first inequality in \eqref{regionI} is saturated.
In Fig.~\ref{fig:regions+},  region I is displayed in blue.

One may wonder that, if we send $N_p\to 0$ in the entropy formula
\eqref{hnes10Apr19}, we can recover the 2-charge entropy
\eqref{S_2-chg}.  However, this does not happen because the regime of
parameters is different; the 3-charge formula \eqref{hnes10Apr19} is
valid only for $N\sim N_p\gg 1$, while the 2-charge formula \eqref{S_2-chg}
is valid only for $N_p=0$.

\subsubsection{Case 3: $z^{\rm bos}_{\ket{-}}$}
\label{sss:z_bos_-}

\begin{figure}[htb]
 \begin{quote}
 \begin{center}
     \includegraphics[height=5cm]{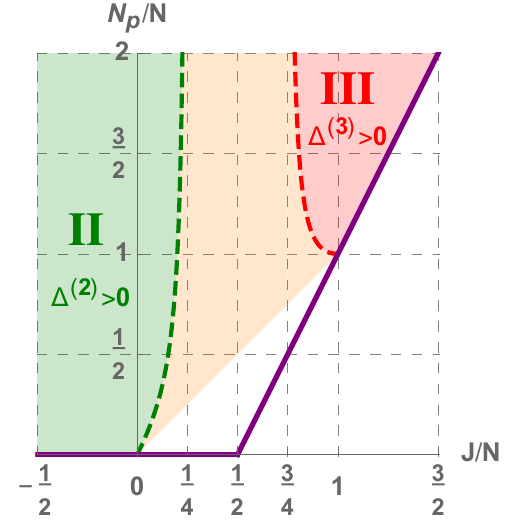} \caption{\sl The regions
  in which states that are counted by $z^{\rm bos}_{\ket{-}}$ exist.
  In the green region, which is given by \eqref{regionII}, the strand
  $\ket{-}_1$ condenses and $\Delta^{(2)}>0$.
  In the red region, which is given by \eqref{regionIII}, the strand
  $(J_{-1}^+)^2\ket{+}_1$ condenses and $\Delta^{(3)}>0$.
  \label{fig:regions-}}
 \end{center}
 \end{quote}
\end{figure}

In this case, we expect that  $\ket{-}_1$ condenses.
The partition function is,
from
\eqref{z_R_bos_a},
\begin{align}
 \log z^{\rm bos}_{\ket{-}}
&=
 \sum_{r=1}^\infty{p^{r} y^{-{r\over 2}} \over r(1-q^r)(1-q^{r}y^{r})}\Biggl[
 {1-2(-1)^r q^r y^{r\over 2}+q^{2r}y^r   \over 1-p^r}
 -
 { q^{3r} y^{2r} (1-2(-1)^r y^{r\over 2}+y^{r})
 \over 1-(p q y)^{r}}
 \Biggr].
\label{itwo10Apr19}
\end{align}
This time, the naive small-$\epsilon$ expansion of the summand and the
results \eqref{zRbos0_ito_a'b'g'} and \eqref{SRbos_0} becomes invalid if
$py^{-1/2}=e^{-\alpha-\gamma/2}=e^{-{1\over
2}(3\alpha'+\beta'-\gamma')}$ is too close to $1$. So, the validity
region of the naive result is
\begin{align}
 3\alpha'+\beta'-\gamma'
 \propto 3N_p'J'+N'J'-N'N_p'
 =-5 J^2 + 2 J N + 4 J N_p - N N_p>0.
\label{iwnv10Apr19}
\end{align}
If the left-hand side becomes very close to zero,
the strand counted by $py^{-1/2}$, namely
$\ket{-}_1$, condenses.
By a computation very similar to the case for $z^{\rm bos}_{\ket{+}}$, we find
\begin{align}
 \log Z\approx -\log{3\alpha'+\beta'-\gamma'\over 2}+{C\over \alpha'\beta'\gamma'}.
\end{align}
Using \eqref{N'Np'J'_a'b'g'}, we get
\begin{align}
 N'  =3\Delta^{(2)}+{C\over \alpha'\beta'\gamma'},\qquad
 N_p'= \Delta^{(2)}+{C\over \alpha'\beta'\gamma'},\qquad
 J'  =-\Delta^{(2)}+{C\over \alpha'\beta'\gamma'},\label{kjsj7May19}
\end{align}
where  now
\begin{align}
 \Delta^{(2)}={1\over 3\alpha'+\beta'-\gamma'}>0.\label{Delta2}
\end{align}
The $\Delta^{(2)}$ terms in \eqref{kjsj7May19}
change $N$ by $2\Delta^{(2)}$ and $J$ by $-\Delta^{(2)}$.  This means
that we have the following condensate:
\begin{align}
 \bigl[\ket{-}_1\bigr]^{2\Delta^{(2)}}.
\end{align}
If we define, just as in \eqref{hdcw10Apr19},
\begin{align}
 N^{(2)}  &=N'  -3\Delta^{(2)}
 %=N  -J-3\Delta_2
 ,\qquad
 N_p^{(2)}=N_p' -\Delta^{(2)}
%=N_p-J -\Delta_2
 ,\qquad
 J^{(2)}  =J'   +\Delta^{(2)}
%=J     +\Delta_2
 ,
\label{jkwr10Apr19}
\end{align}
then $\alpha',\beta',\gamma'$ are given by \eqref{hdbs10Apr19}, with the
superscript ``(1)'' replaced by ``(2)''.  Again, $\Delta^{(2)}$ is
determined by the condition $3\alpha'+\beta'-\gamma'\approx 0$, which
amounts to
\begin{align}
 3 N_p^{(2)} J^{(2)} + N^{(2)} J^{(2)} - N^{(2)} N_p^{(2)}=0.
\end{align}
This gives
\begin{align}
\Delta^{(2)}=
\frac{1}{9} \left(-7 J+N+3 N_p-\sqrt{4 J^2+4 J N+N^2-6 J N_p-3 N N_p+9N_p^2}\right).
\label{jsbx10Apr19}
\end{align}
A choice for the sign has been made just as for the case with $z^{\rm
bos}_{\ket{+}}$. 
The region in which $\Delta^{(2)}\ge 0$ is in the
complement of the region given by~\eqref{iwnv10Apr19},
namely,
\begin{align}
 N_p\ge {J (5 J - 2 N)\over 4(J - N/4)},\qquad
 N_p\ge 0,\qquad
 -{N\over 2}\le J\le {N\over 4}\qquad 
 \text{(region II)}.
 \label{regionII}
\end{align}
Here, we also included inequalities that come from the condition
$N^{(2)},N_p^{(2)},J^{(2)}\ge 0$.  
In Fig.~\ref{fig:regions-},  region II is displayed in green.
The entropy in this region is given by the same expression as before,
Eq.~\eqref{hnes10Apr19}, with ``(1)'' replaced by~``(2)''.

\bigskip
We just found that condensation occurs in the region with $J\in[-{N\over
2},{N\over 4}]$. Recall that we have spectral flow symmetry \eqref{spfl}
and the charge-flipping symmetry $j\to -j$. Combining these, we can show
that there is symmetry that maps states with $J\in[-{N\over 2},{N\over
4}]$ into those with $J\in[{3N\over 4},{3N\over 2}]$.  More precisely,
flowing from the R sector to the NS sector, flipping the sign of $j$ there,
and then flowing back to the R sector, we have the following map on 
a strand of length $k$:
\begin{align}
 h\to h-2j+k,\qquad
 j\to k-j.\label{mqiq7May19}
\end{align}
Under this map, $\ket{-}_1$ with $h={1\over 4},j=-{1\over 2}$ goes to
$(J_{-1}^{+})^2\ket{-}_{1}$ with $h={9\over 4},j={3\over 2}$.  This
means that, there must be a region with $J\in[{3N\over 4},{3N\over 2}]$
where $(J_{-1}^{+})^2\ket{-}_{1}$ condenses.\footnote{In the NS sector,
$\ket{-}_1^{\rm R}$ corresponds to a chiral primary $\ket{+}_1^{\rm NS}$
with $h^{\rm NS}=j^{\rm NS}=1$, while $(J_{-1}^{+})^2\ket{-}_{1}^{\rm
R}$ corresponds to an anti-chiral primary $(J_0^-)^2\ket{+}_1^{\rm NS}$
with $h^{\rm NS}=-j^{\rm NS}=1$.  }

The map \eqref{mqiq7May19} can be realized in the partition function by the
replacement \begin{align} p\to pqy,\qquad q\to q,\qquad y\to
q^{-2}y^{-1}.
\end{align}
It is straightforward to show that \eqref{itwo10Apr19} is invariant
under this.  In particular, the very first term involving $py^{-1/2}$,
which led to condensation of $\ket{-}_1$, is mapped into $p q^2 y^{3/2}=
e^{-{1\over 2}(-\alpha'+\beta'+3\gamma')}$.  When this is very close to
one, the partition function is 
\begin{align}
 \log z^{\rm bos}_{\ket{-}}\approx -\log{-\alpha'+\beta'+3\gamma'\over 2}+{C\over \alpha'\beta'\gamma'}.
\end{align}
So, this time,
\begin{align}
 N'  =-\Delta^{(3)}+{C\over \alpha'\beta'\gamma'},\qquad
 N_p'= \Delta^{(3)}+{C\over \alpha'\beta'\gamma'},\qquad
 J'  =3\Delta^{(3)}+{C\over \alpha'\beta'\gamma'},\label{mvpg7May19}
\end{align}
where  now
\begin{align}
 \Delta^{(3)}={1\over -\alpha'+\beta'+3\gamma'}>0.
\end{align}
The $\Delta^{(3)}$ terms in \eqref{mvpg7May19} change $N$ by
$2\Delta^{(3)}$, $N_p$ by $4\Delta^{(3)}$, and $J$ by $3\Delta^{(3)}$.
This means that we have the following condensate:
\begin{align}
 \bigl[(J_{-1}^+)^2\ket{-}_1\bigr]^{2\Delta^{(3)}}.
\end{align}
If we define
\begin{align}
 N  ^{(3)}&=N'   +\Delta^{(3)}
%= N  -J +\Delta^{(3)} 
 ,\qquad
 N_p^{(3)}=N_p' -\Delta^{(3)} 
 %= N_p-J -\Delta^{(3)} 
 ,\qquad
 J  ^{(3)}=J'  -3\Delta^{(3)}
%= J    -3\Delta^{(3)} 
,
\label{jkrr10Apr19}
\end{align}
then $\alpha',\beta',\gamma'$ are given by \eqref{hdbs10Apr19}, with the
superscript ``(1)'' replaced by ``(3)''.   $\Delta^{(3)}$ is
determined by the condition $-\alpha'+\beta'+3\gamma'\approx 0$, which
amounts to
\begin{align}
 -N_p^{(3)} J^{(3)} + N^{(3)} J^{(3)} +3 N^{(3)} N_p^{(3)}=0.
\end{align}
This gives
\begin{align}
\Delta^{(3)}=
\frac{1}{9} \left(J-3 N+3 N_p-\sqrt{28 J^2-6 J (4 N+5 N_p)+9 \left(N^2+N
   N_p+N_p^2\right)}\right).
 \label{jsll10Apr19}
\end{align}
The region in which $\Delta^{(3)}\ge 0$ is given by
\begin{align}
 N_p\ge {J (3 J - 2 N)\over 4(J - 3 N/4)},\qquad 
 N_p>2J-N,\qquad
 {3N\over 4}\le J\le {3N\over 2}\qquad 
 \text{(region III)}.
 \label{regionIII}
\end{align}
Here, we also included inequalities that come from the condition
$N^{(3)},N_p^{(3)},J^{(3)}\ge 0$.  
In Fig.~\ref{fig:regions-},  region III is displayed in red.
The entropy in this region is given by 
Eq.~\eqref{hnes10Apr19}, with ``(1)'' replaced by~``(3)''.

\subsubsection{Case 4: $z^{\rm fer}_{\ket{a}}$}
\label{sss:z_fer}

Lastly let us consider the fermionic case.  Because the terms that were
dangerous for the bosonic case, namely the very first term and the very
last term in \eqref{z_R_bos_a}, come with alternating signs $(-1)^r$ in
the fermionic partition function \eqref{z_R_fer_a}, these terms do not
lead to divergences.  This means that the naive small-$\epsilon$
expansion of the summand of the fermionic partition
function~\eqref{z_R_fer_a} is always justified and we obtain
\begin{align}
 \log z^{\rm fer}_{\ket{a}}
 \approx{C\over \alpha' \beta' \gamma'},
\end{align}
independent of the value of $a$.  Note that this is identical to the
``bosonic'' result, \eqref{zRbos0_ito_a'b'g'}, without $1\over 2$ that
one might have expected for a ``fermionic'' partition function.  This is
because the descendant states we are counting always include both
bosonic and fermionic states, whether the R ground state on which these
states are built is bosonic or fermionic.  Namely, if $\ket{\psi}^{\rm
R}_k$ is bosonic (fermionic), the first and third lines of
\eqref{1p_R_states} are bosonic (fermionic) and the second line of
\eqref{1p_R_states} are fermionic (bosonic).

\subsection{Entropy for the full partition function}
\label{ss:full_pf}

Let us put together the pieces obtained above and estimate the entropy
for the full partition function \eqref{ZT4,ZK3_as_sum}.  A naive
small-$\epsilon$ expansion of the summand gives
\begin{align}
 \log Z_{T^4,\rm K3}\approx
 {c\,C\over \alpha'\beta'\gamma'},\label{mrxw8May19}
\end{align}
where $C$ was defined in \eqref{zRbos0_ito_a'b'g'} and
\begin{align}
 c=\begin{cases}
    16 & T^4,\\
    24 & K3.
   \end{cases}
\end{align}
Just as for $\log z^{\rm bos}_{\ket{0}}$ discussed in section
\ref{sss:z_bos_0}, the entropy is computed to be
\begin{align}
 S=4(c\, C N' N_p' J')^{1/4}=
 4\left[c\,CJ(N-J)(N_p-J)\right]^{1/4}\equiv S_0,
 \label{haxa8May19}
\end{align}
where  $N',N_p',J'$ were defined in \eqref{def_N'Np'J'}.
However, this expression is not valid for all values of $N_p,J$, due
to condensation of certain length-one strands.

In region I defined by \eqref{regionI}, $\alpha'-\beta'+\gamma'\to +0$
(more precisely, the left-hand side becomes~$\cO(\epsilon^4)$) due to the condensation of the strand~$\ket{+\pm}_1$.
The partition function in this case is modified from \eqref{mrxw8May19}
as
\begin{align}
 \log Z\approx
 -2\log{\alpha'-\beta'+\gamma'\over 2}
 +{c\,C\over \alpha'\beta'\gamma'}.
\end{align}
The condensate has the following form:
\begin{align}
 \bigl[\ket{++}_1\bigr]^x 
 \bigl[\ket{+-}_1\bigr]^y,\qquad
 x+y = 2\Delta^{(1)},
\end{align}
where we defined
\begin{align}
 \Delta^{(1)}\equiv {2\over \alpha'-\beta'+\gamma'}
\end{align}
(the factor $2$ as compared to \eqref{Delta1} is due to the 2 in front
of $\log z^{\rm bos}_{\ket{+}}$ in \eqref{ZT4,ZK3_as_sum}).
The entropy in region I is
\begin{align}
 S_{\rm I}&=4\bigl(c\, C N^{(1)} N_p^{(1)} J^{(1)}\bigr)^{1/4}\notag\\[1ex]
 &=4\biggl[{2c\,C\over 27} \biggl((-{N\over 2} + J - N_p) (-2N + 4 J - N_p) (-N + 2 J + N_p) \notag\\[-2ex]
 &\hspace{20ex}
 + (N^2 + 4 J^2 +
      N N_p + N_p^2 - 2 J (2N + N_p))^{3/2}\biggr)\biggr]^{1/4}
,\label{hpvj13May19}
\end{align}
where
$N^{(1)},N_p^{(1)},J^{(1)}$ are defined by \eqref{hdcw10Apr19}
and $\Delta^{(1)}$ is given by \eqref{jrrv10Apr19}.

In region II defined by \eqref{regionII}, $3\alpha'+\beta'-\gamma'\to
+0$ due to the condensation of the strand~$\ket{-\pm}_1$.  The partition
function in this case is
\begin{align}
 \log Z\approx
 -2\log{3\alpha'+\beta'-\gamma'\over 2}
 +{c\,C\over \alpha'\beta'\gamma'}.
\end{align}
The condensate is
\begin{align}
 \bigl[\ket{-+}_1\bigr]^x 
 \bigl[\ket{--}_1\bigr]^y,\qquad
 x+y = 2\Delta^{(2)},\qquad
 \Delta^{(2)}\equiv {2\over 3\alpha'+\beta'-\gamma'}.
\end{align}
The entropy in region II is
\begin{align}
 S_{\rm II}&=4\bigl(c\, C N^{(2)} N_p^{(2)} J^{(2)}\bigr)^{1/4},\notag\\[1ex]
 &=4\biggl[{2c\,C\over 243} 
\biggl(-({N\over 2} + J - 3 N_p) (2N + 4 J - 3 N_p) (N + 2 J + 3 N_p) 
\notag\\[-2ex]
 &\hspace{20ex}
 + (N^2 + 4 J^2 + 
    J (4 N- 6 N_p) - 3 N N_p + 9 N_p^2)^{3/2}\biggr)\biggr]^{1/4}
\label{hpvl13May19}
\end{align}
where
$N^{(2)},N_p^{(2)},J^{(2)}$ are defined by \eqref{jkwr10Apr19}
and $\Delta^{(2)}$ is given by \eqref{jsbx10Apr19}.

In region III defined by \eqref{regionIII}, $-\alpha'+\beta'+3\gamma'\to
+0$ due to the condensation of the strand~$(J_{-1}^+)^2\ket{-\pm}_1$.
The partition function in this case is
\begin{align}
 \log Z\approx
 -2\log{-\alpha'+\beta'+3\gamma'\over 2}
 +{c\,C\over \alpha'\beta'\gamma'}.
\end{align}
The condensate is
\begin{align}
 \Bigl[(J_{-1}^+)^2\ket{-+}_1\Bigr]^x
 \Bigl[(J_{-1}^+)^2\ket{--}_1\Bigr]^y,\qquad
 x+y = 2\Delta^{(3)},\qquad
 \Delta^{(3)}\equiv {2\over -\alpha'+\beta'+3\gamma'}.
\end{align}
The entropy in region III is
\begin{align}
 S_{\rm III}&=
 4\bigl(c\, C N^{(3)} N_p^{(3)} J^{(3)}\bigr)^{1/4}\notag\\[1ex]
 &=4\biggl[{2c\,C\over 243} \biggl((-{3N\over 2} + 5 J - 3 N_p) (3N + 2 J - 3 N_p) (-6N + 8 J - 
      3 N_p) \notag\\[-2ex]
 &\hspace{20ex}
 + (28 J^2 - 6 J (4N + 5 N_p) + 9 (N^2 + N N_p + N_p^2))^{3/2}\biggr)\biggr]^{1/4}
 ,\label{hpvp13May19}
\end{align}
where
$N^{(3)},N_p^{(3)},J^{(3)}$ are defined by \eqref{jkrr10Apr19}
and $\Delta^{(3)}$ is given by \eqref{jsll10Apr19}.  

\begin{figure}[htb]
  \begin{quote}
 \begin{center}
  \includegraphics[height=8cm]{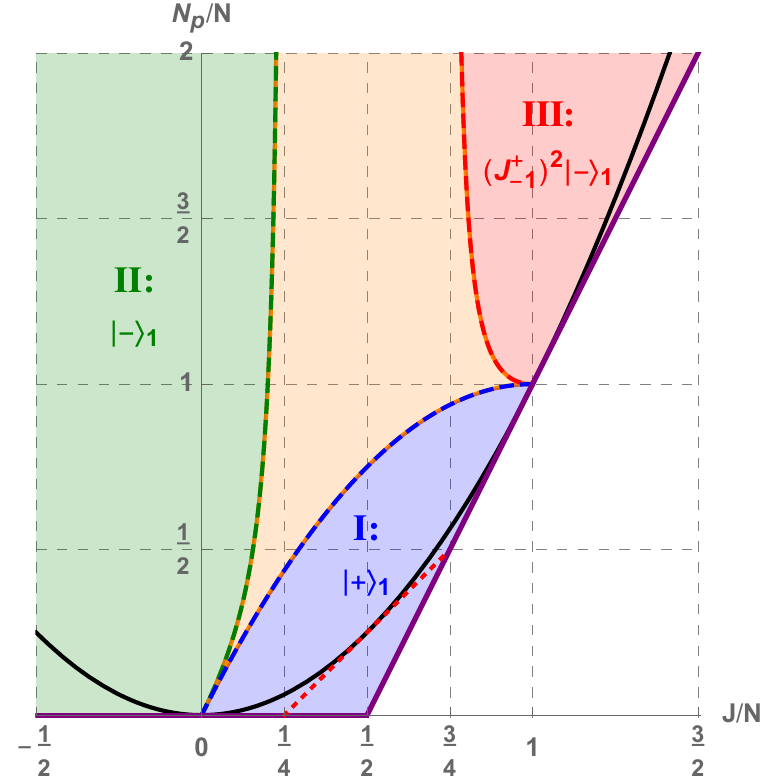} % Math'ca notes on 2019.04.22
  \caption{\sl The phase
   diagram on the $J$-$N_p$ plane.  $N_p=L_0-{N\over 4}$.
  In the orange region, no condensation happens and the entropy is given
  by \eqref{haxa8May19}.
  In region~I (light blue shaded), condensate of $\ket{+}_1$ forms
  and $\Delta^{(1)}>0$.  
  In region~II (light green shaded), condensate of $\ket{-}_1$ forms
 $\Delta^{(2)}>0$.
  In region~III (light red shaded), condensate of
  $(J_{-1}^+)^2\ket{-}_1$ forms and $\Delta^{(3)}> 0$.
  Above
   the solid black line $N_p=J^2$, single-center black holes exist.  The
   purple solid lines represent the unitarity bound below which 
   no states exist.
  \label{fig:regions}}.
\end{center}
  \end{quote} 
\end{figure}

\begin{figure}[htb]
  \begin{quote}
 \begin{center}
  \vspace*{-4ex}
   \includegraphics[height=8cm]{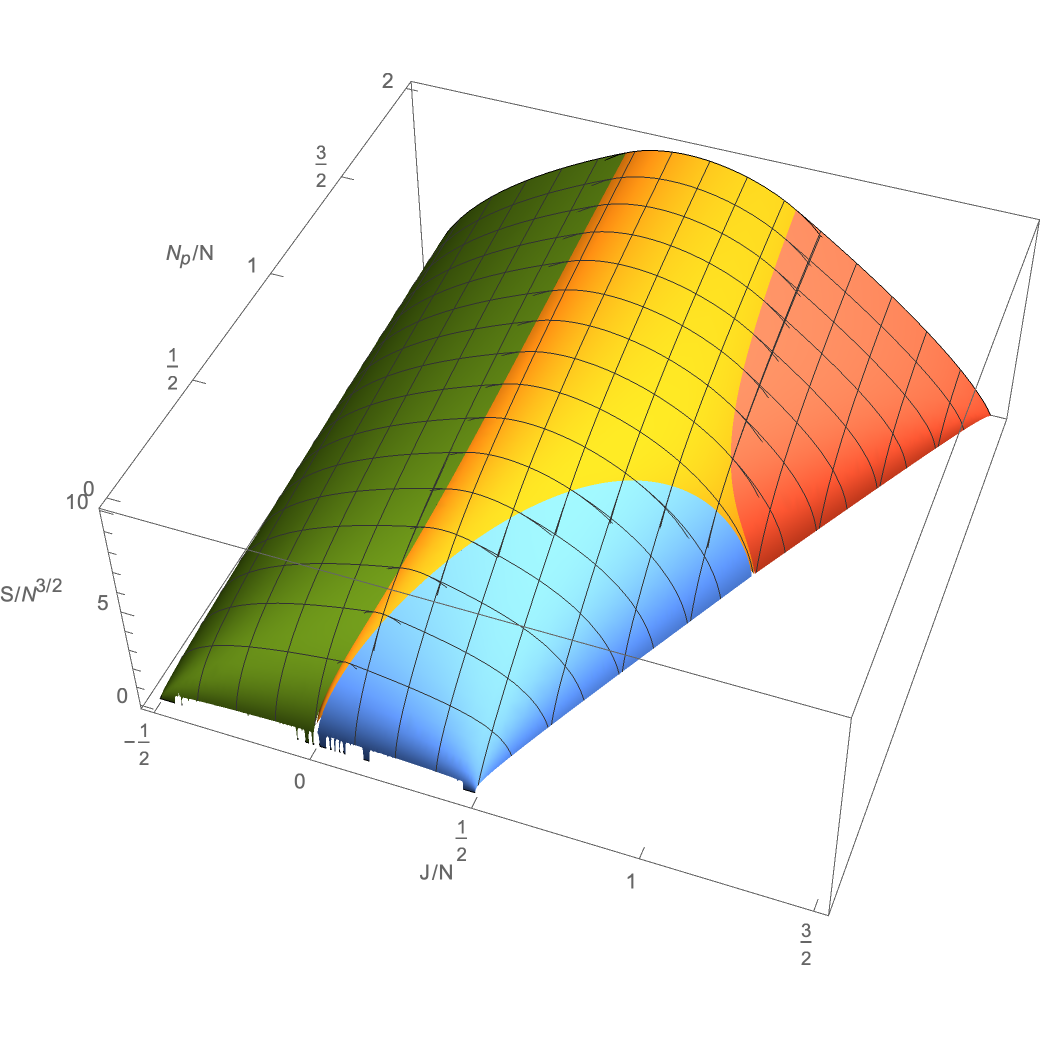}
  \vspace*{-4ex}
\caption{\sl The plot of the
  entropy for K3 ($c=24$), obtained by patching together the formulae
  \eqref{haxa8May19}, \eqref{hpvj13May19}, \eqref{hpvl13May19}, and
  \eqref{hpvp13May19} in different regions.  See Fig.~\ref{fig:regions}
  for which color corresponds to which region.  The case with $T^4$ is
  identical, except for the overall factor.
  \label{fig:S_K3}}
\end{center}
  \end{quote} 
\end{figure}

In Fig.~\ref{fig:regions}, different regions on the $J$-$N_p$ plane are
displayed.  In regions I--III, certain condensate forms and one must use
different formulas (Eqs.~\eqref{haxa8May19}, \eqref{hpvj13May19},
\eqref{hpvl13May19}, and \eqref{hpvp13May19}).
In Fig.~\ref{fig:S_K3}, a plot of the entropy is given, for the case
with $M_4={\rm K3}$.  As one can see, the entropy as a function of
$N_p,J$ is smooth across the boundary of different regions.  The $T^4$
case is identical, except for the overall factor due to difference in
the value of $c$.

The general formulas for entropy are complicated, but for some special
cases they simplify.  For large $N_p\gg N,|J|$,\footnote{ See footnote
\ref{ftnt:Np>>N>>>1}.} region I becomes irrelevant while regions II and
III become vertical strips on the $J$-$N_p$ plane.  The entropy for this
regime is given by
\begin{align}
 S=
\begin{cases}
 S_{\rm II}\approx 4\sqrt{J+{N\over  2}}\,\left({c\,C\over 3}N_p\right)^{1/4} &(-{N\over 2}\le J\le {N\over 4})\\[2ex]
 S_0\approx 4[c\,CJ(N-J)N_p]^{1/4}& ({N\over 4}\le J\le {3N\over 4})\\[2ex]
 S_{\rm III}\approx 4\sqrt{{3N\over 2}-J}\,\left({c\,C\over 3}N_p\right)^{1/4}&({3N\over 4}\le J \le {3N\over 2})\\
\end{cases}
\label{ghqd14May19}
\end{align}

One can also consider the entropy for $J=0$.  This is entirely in region
II and the entropy is
\begin{align}
 S|_{J=0}&=S_{\rm II}|_{J=0}\notag\\
 &=4\biggl[
{2c\,C\over 243} \biggl(
-({N\over 2} - 3 N_p) (2 N - 3 N_p) (N + 3 N_p)
 +(N^2-3N N_p+9N_p^2)^{3/2} \biggr)\biggr]^{1/4}.
\end{align}
This behaves for small and large $N_p$ as\footnote{Because we have
already taking the large $N,N_p$ limit, this means that $1\lll N_p\ll N$
and $N_p\gg N\ggg 1$.}
\begin{align}
 S|_{J=0}\approx
 \begin{cases}
  \displaystyle
 \frac{2^{3/4} \pi c^{1/4} }{3^{1/4}}N^{1/4}N_p^{1/2}
  &\qquad (N_p\ll N)\\[2ex]
  \displaystyle
 \frac{2^{3/4} \pi c^{1/4}}{\sqrt{3}}N^{1/2}N_p^{1/4}
  &\qquad (N_p\gg N)\\
 \end{cases}
\end{align}

\subsection{Comparison with the black-hole entropy}

In the above, we derived the entropy $S$ of the ensemble of
supergravitons, or equivalently, superstrata.  Although the precise
functional form depends on the region on the $J$-$N_p$ plane, we
universally have
\begin{align}
 S= \cO(N^{3/4}),
 \qquad{\rm for}\qquad
 N\gg 1,\quad N_p, J=\cO(N).\label{jhtd13May19}
\end{align}
This is parametrically smaller than the black-hole entropy,
\begin{align}
 S=2\pi \sqrt{N N_p-J^2}=\cO( N).
\end{align}
Namely, as expected, superstrata obtained by non-linear completion of
supergraviton gas states around empty AdS$_3\times S^3$ have much less
entropy than the black hole.

In \eqref{ghqd14May19}, we estimated the entropy for $N_p\gg N\sim |J|$.
In that case, we obtain
\begin{align}
 S\sim N^{1/2} N_p^{1/4},\label{gykg14May19}
\end{align}
while, in the same regime,
\begin{align}
 S_{\rm BH}\sim N^{1/2}N_p^{1/2}.
\end{align}
So, the superstrata are too few because of the smaller power of $N_p$.

%The power of $N$ and $N_p$ in \eqref{gykg14May19} can be understood as
%follows.  For simplicity, consider a simpler ensemble, which is a gas of
%the particles:
%\begin{align}
% L_{-1}^n (J_0^+)^l \ket{0}_k,\qquad
% k\ge 1,\quad 0\le l\le  k,\quad n\ge 0.
%\end{align}
%If we do not count the angular momentum, the partition function is
%\begin{align}
% Z=\prod_{k,l,n}{1\over 1-p^k q^n}.
%\end{align}
%We can estimate this partition function as before as
%\begin{align}
% \log Z
% &=-\sum_{k,l,n} \log(1-p^{k}q^{n})
% =\sum_{k,l,n,r} {p^{kr}q^{nr}\over r}
% =\sum_{k,n,r} { k p^{kr}q^{nr}\over r}
% \notag\\
% &=\sum_{r} { p^r\over r(1-p^r)^2(1-q^r)}
% \approx \sum_{r} { 1\over r^4 \alpha^2 \beta}
% ={C\over \alpha^2 \beta}.
%\end{align}
%From this, it is straightforward to obtain
%\begin{align}
% \alpha \sim N^{-1/2}N_p^{1/4},\qquad \beta \sim N^{1/2}N_p^{-3/4}
%\end{align}
%and the entropy is
%\begin{align}
% S\sim N^{1/2} N_p^{1/4}.
%\end{align}
%The power of $N$ comes from the proliferation of ``species'' by the
%action of $J_{-1}$

In \cite{Bena:2010gg}, the entropy of the possible microstate geometries
in the D1-D5 system was estimated based on the entropy enhancement
mechanism.  This mechanism says that the entropy carried by supertubes
is strongly enhanced by putting them in the throat of a smooth
multi-center solution \cite{Bena:2005va, Berglund:2005vb}.  Because
supertubes can become smooth geometries upon backreaction
\cite{Lunin:2001jy, Lunin:2002iz}, the entire configuration is expected
to become a smooth geometry. This mechanism is expected to lead to
microstate geometries with a large entropy.  The estimate of
\cite[Eq.~(6.18)]{Bena:2010gg} is that, if $N_1 \sim N_5\sim N_p\sim Q$
then the entropy is $S\sim Q^{5/4}$.  This agrees with
\eqref{gykg14May19} if we set $N\sim Q^2,N_p\sim Q$.  However, the
scaling regime appear to be different and it is not clear if this is a
sensible comparison.  We leave for future research a further
investigation into this interesting issue.

\subsection{Spectral flows}
\label{ss:spectral_flows}

In the above, we considered the states of the form \eqref{1/8_sugrtn_R},
which are within the range $J\in [-{1\over 2}N,{3\over 2}N]\equiv I_0$.
However, by spectral flow \eqref{spfl}, we can map these states into
other range.  For example, by spectral flow with $\eta=1$, we obtain
states that sit within the range $J\in [{1\over 2}N, {5\over 2}N]\equiv
I_1$.  More generally, the states spectral-flowed by parameter
$\eta\in\bbZ$ sit in the range $[(\eta-{1\over 2})N,(\eta+{3\over
2})N]\equiv I_\eta$.  These states do not describe fluctuations around
empty AdS$_3\times S^3$ but are fluctuations around spectral flows of
empty AdS$_3\times S^3$.  However, the corresponding superstrata
solutions do represent microstate geometries as the original (unflowed)
superstrata do and their entropy must also be taken into account.
The entropy of the spectral-flowed states can be easily obtained by
applying the spectral flow \eqref{spfl} to the entropy formulas obtained
in section \ref{ss:full_pf}.
Let us discuss how the phase diagram in Fig.~\ref{fig:regions} changes
if we consider such spectral-flowed states.

The interval $I_\eta$ overlaps with $I_{\eta+1}$ in the range
$[(\eta+{1\over 2})N,(\eta+{3\over 2})N]$. In the overlapping region,
one can show that the entropy of the $\eta$-flowed states is dominant in
$[(\eta+{1\over 2})N,(\eta+1) N]$ while the entropy of the
$(\eta+1)$-flowed states is dominant in $[(\eta+1) N,(\eta+{3\over
2})N]$.  This means that the $\eta$-flowed states are dominant for $J\in
[\eta N,(\eta +1)N]$.  Actually, if we take the $J\to -J$ symmetry also
into account, the boundary between different regions of dominance can
not be anywhere else than $J=nN$, $n\in\bbZ$.

\begin{figure}[htb]
  \begin{quote}
 \begin{center}
  \vspace*{-6ex}
  \includegraphics[height=6cm]{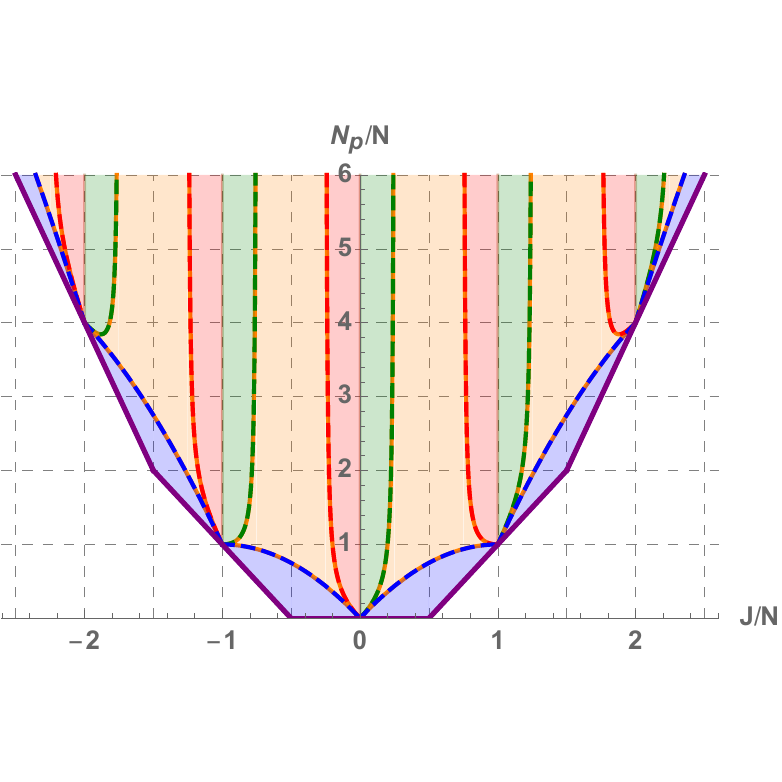} % Math'ca notes on 2019.05.14
  \vspace*{-6ex}
  \caption{\sl The phase diagram on the $J$-$N_p$ plane, now taking into
   account of the spectral flowed states.
  \label{fig:flowed_regions}}.
\end{center}
  \end{quote} 
\end{figure}

The regions of dominance on the $J$-$N_p$ plane, taking into account
flowed states, are shown in Fig.~\ref{fig:flowed_regions}.  The meaning
of the colors is the same as that in Fig.~\ref{fig:regions}.  In the
flows of regions I, II, and III (blue, green, and red), the
flows of the states $\ket{+}_1$, $\ket{-}_1$, and
$(J_{-1}^+)^2\ket{-}_1$ condense, respectively.

\section{Discussion}
\label{sec:discussion}

In this paper, we evaluated the partition function of CFT states that
correspond to superstrata in the bulk.  These superstrata can be thought
of as non-linear completion of supergravitons around empty AdS$_3\times
S^3$, and represent a certain class of microstates of the D1-D5-P black
hole.  We found that the entropy computed from the partition function is
parametrically smaller than the entropy of the black hole with the same
charges.  Therefore, these superstrata based on ${\rm AdS}_3\times S^3$
are not typical microstates of the black hole.

Our result is similar to \cite{deBoer:2009un}, in which gravity
microstates (multi-center solutions) of a certain 4-charge system were
counted and the entropy was found to be parametrically smaller than the
corresponding black-hole entropy.  Instead, the entropy was found to be
equal to that of the supergraviton gas in an AdS$_3$ background.
However, what is different in our setup compared to that in
\cite{deBoer:2009un} is that we have a boundary CFT which is in a better
theoretical control and gives us hints as to what kind of state we are
missing.
The ingredients that the superstrata counted in this paper lack are
higher and fractional modes.  The superstrata constructed in
\cite{Bena:2016agb} involve some fractional modes because they are not
based on ${\rm AdS}_3\times S^3$ but on the orbifold $({\rm AdS}_3\times
S^3)/\bbZ_k$.  It would be interesting to generalize the counting of the
current paper to include such superstrata.\footnote{One would have to be
careful to the fact that some supergravitons on $({\rm AdS}_3\times
S^3)/\bbZ_k$ with different values of~$k$ actually represent the same
state.}  However, the class of fractional modes that these superstrata
\cite{Bena:2016agb} involve are restricted and we must look for the bulk
realization of states involving more general fractional modes and also
higher modes.

At the orbifold point of the D1-D5 CFT, it is clear how to construct BPS
states that involve higher and fractional modes.  On a strand of length
$k\ge 1$, we have modes such as $L_{-{n\over k}}, G^{\alpha
A}_{-{n+1/2\over k}}, J^i_{-{n\over k}}$ with $n\ge 1$ (in the NS
sector) and we are free to excite them as long as the total $L_0$ on the
strand is an integer.  If we go away from the orbifold point, many of
these states are known to lift and become non-BPS \cite{Gava:2002xb,
Guo:2019pzk}.  The strand excited by generators with general mode
numbers is conjectured to correspond to a string propagating in ${\rm
AdS}_3\times S^3$ with string oscillators excited on it \cite{Lunin:2002fw,
Gomis:2002qi, Gava:2002xb} (see also \cite{David:2008yk}).
On the other hand, the modified elliptic genus for $T^4$ computed from
CFT gets contribution only from ``identical-strand'' states, namely, one
must have $N\over k$ length-$k$ strands with identical excitations on
every one of them \cite{Maldacena:1999bp}.  This suggests that, a single
string with oscillators excited on it is non-BPS away from the orbifold
point but, if we have as many copies of the same string as is allowed by
the stringy exclusion principle, they give rise to a BPS state because
the binding energy cancels the excitation energy of the string
oscillators (this statement was confirmed by an explicit CFT computation
for a particular excited strand in \cite{Guo:2019pzk}).

This seems to suggest that states involving general (higher and
fractional) modes are represented in gravity picture by non-BPS strings
propagating in some background.  However, that is a picture valid for a
small number of strings.  If we have $\cO(N)$ such strings, all in the
same state, they are likely to have an alternative description in terms
some puffed-out branes or backreacted geometries, by polarization due to
Myers' effect \cite{Myers:1999ps} or the supertube effect
\cite{Mateos:2001qs}.\footnote{Examples of such phenomena are common in
string theory, including giant gravitons \cite{McGreevy:2000cw}, which
are gravitons polarizing into D-branes, or Wilson loops
\cite{Drukker:2005kx}, which are fundamental strings on top of each
other, being better described in terms of D-branes.  Those D-brane
configurations have a gravity description in terms of smooth geometries
\cite{Lin:2004nb, Yamaguchi:2006te}.}
Such brane configurations or microstate geometries may still be non-BPS,
as the original (unpuffed-out) string, but it is logically possible that they
actually become BPS in this alternative description in certain
situations. Indeed, recall that, it was observed \cite{Bena:2011zw} that
there is some correspondence between states that are BPS at the orbifold
point of the D1-D5 CFT but become non-BPS away from the orbifold point,
and solutions in supergravity that are BPS when moduli are trivial
(internal NS-NS and RR fields vanish) but become non-BPS when generic
moduli are turned on \cite{Dabholkar:2009dq, Bossard:2019ajg}.  Because
these string states are BPS at the orbifold point, it is conceivable
that the puffed-out strings are represented by some BPS configurations
in supergravity when moduli are trivial.  In this view, it would be
interesting to study possible relations between such puffed-out branes
and known BPS brane configurations in the D1-D5 system
\cite{Mandal:2007ug}.

Another interesting background to look at is ${\rm AdS}_2$.  It has been
argued that the bulk microstates of BPS black holes live in the close
vicinity of the horizon and are represented by asymptotically AdS$_2$
configurations with vanishing angular momentum \cite{Sen:2009vz,
Chowdhury:2014yca, Chowdhury:2015gbk}.  In the language of quiver
quantum mechanics, they correspond to states in the so-called pure Higgs
branch \cite{Bena:2012hf}.
Recently, some pieces of evidence have been found for the
relevance of microstate geometries with AdS$_2$ asymptotics for
black-hole microstates.
In \cite{Fernandez-Melgarejo:2017dme}, certain configurations of
codimension-two branes were explicitly constructed as candidates for
black-hole microstates.  Interestingly, such solutions can exist only
with AdS$_2\times S^2$ asymptotics, due to the monodromic structure of
the harmonic functions required of the codimension-two branes. These
states have vanishing angular momentum due to a cancellation mechanism
by interaction between branes.
In a more recent paper \cite{Heidmann:2018vky}, an exhaustive search for
smooth multi-center solutions of type \cite{Bena:2005va,
Berglund:2005vb} with minimum possible charges was carried out and, it
was found that the bubble equations allow exactly as many solutions as
predicted by the quiver quantum mechanics living on the branes 
\cite{Chowdhury:2014yca, Chowdhury:2015gbk}.
These solutions have no angular momentum and are all asymptotically
AdS$_2\times S^2$, which appears to suggest that states in the pure
Higgs branch are represented by asymptotically AdS$_2$ configurations in
the bulk.
Therefore, it would be highly interesting to study gravity microstates
with D1-D5-P charges (superstrata or any other configurations, with or
without supergravity description) imposing a strict AdS$_2$ boundary
condition at infinity.  Having no AdS$_3$ region makes it difficult to
identify the corresponding dual state in the D1-D5 CFT, but considering
an AdS$_2$ limit of superstrata solutions with known CFT duals, such as
\cite{Bena:2018bbd}, can be useful for that purpose.

\begin{figure}[htb]
  \begin{quote}
 \begin{center}
\begin{tabular}{cc}
    \includegraphics[height=7cm]{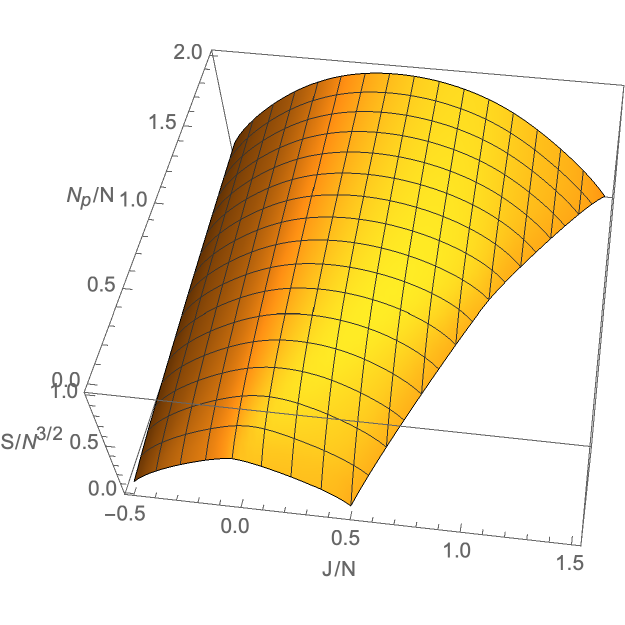}
 &
    \includegraphics[height=7cm]{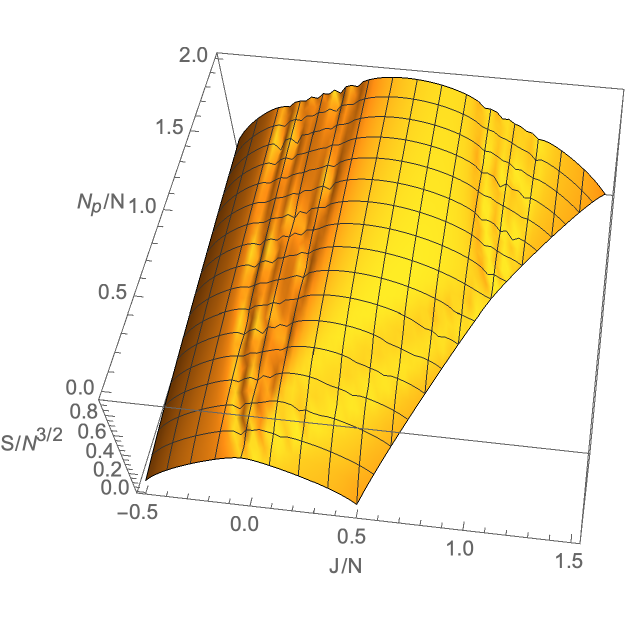}
 \\
 (a) & (b)
\end{tabular}
\caption{\sl The plot of the growth of the coefficients of (a)
  supergraviton partition function and (b) supergraviton elliptic genus
  for K3 computed numerically for $N=16$.  What is plotted is $S\equiv
  \log|c(N,N_p,J)|$ where we expand partition function and elliptic
  genus as $Z_N=\sum_{N_p,J}c(N,N_p,J)q^{N_p}y^J$. \label{fig:S_K3_num}}
\end{center}
  \end{quote} 
\end{figure}

In the current paper, we computed the partition function of
supergravitons (or equivalently, superstrata).  Another very interesting
quantity to investigate is the (modified) elliptic genus.
It was shown by de Boer \cite{deBoer:1998us} that the elliptic genus
computed from CFT agrees with the elliptic genus computed by enumerating
(with signs) supergravitons, for $L_0^{\rm NS}\le {N+1\over 4}$ for K3
(for the modified elliptic genus for $T^4$, the bound is $L_0^{\rm NS}<
{N\over 4}$ \cite{Maldacena:1999bp}).  This de Boer bound is shown as
red dashed lines in Fig.~\ref{fig:JNp_plane_R} and other Figures.  Above
this bound, new primary states appear, which are responsible for the
black-hole entropy.  Above this bound and below the black-hole bound,
$L_0^{\rm NS}=J^2/N+N/4$, the CFT elliptic genus is different from the
supergraviton elliptic genus but does not yet show a black-hole growth.
It is quite interesting to study the behavior of the elliptic genera in
this intermediate region to understand the nature of states that are not
captured by supergravitons.
The supergraviton elliptic genus is obtained simply by changing some
signs in partition functions such as \eqref{z_NS_bos}.  However, because
the coefficients of elliptic genus can be positive or negative, unlike
partition function, we cannot use a thermodynamic approximation we used
in this paper to estimate its growth.  We probably need a more
sophisticated way to rewrite the elliptic genus to be able to accurately
estimate it, such as the one in \cite{Benini:2018ywd} or its relation to
four-dimensional indices \cite{Castro:2008ys}.
In Fig.~\ref{fig:S_K3_num}, we plotted the supergraviton partition
function and the supergraviton elliptic genus for K3 for $N=16$.  We
can see that the elliptic genus has structure more non-trivial than the
partition function.  If would be interesting to understand this
structure.

%\vspace{3cm}
%%%%%%%%%%%%%%%%%%%%%%%%%%%%%%%%%%%%%
\section*{Acknowledgments}
%%%%%%%%%%%%%%%%%%%%%%%%%%%%%%%%%%%%%

\vspace{-2mm}

I thank Iosif Bena, Jan de Boer, Pierre Heidmann, Stefano Giusto, Emil
Martinec, David Turton, Rodolfo Russo and Nick Warner for useful
discussions.
The work of MS was supported in part by JSPS KAKENHI Grant Numbers
16H03979, and MEXT KAKENHI Grant Numbers 17H06357 and 17H06359.

\appendix

\section{The NS sector}
\label{app:NS}

In the main text, we studied the entropy of the R states of the form
\eqref{1/8_sugrtn_R}, because bulk states naturally live in the R
sector.  In this Appendix, we discuss the NS sector formulas, which are
obtained by applying the map \eqref{spfl+flip} to the R formulas.  Some
formulas are simpler in the NS sector because of the $J\leftrightarrow
-J$ symmetry.

\begin{figure}[htb]
  \begin{quote}
 \begin{center}
  \vspace*{-8ex}
  \includegraphics[height=8cm]{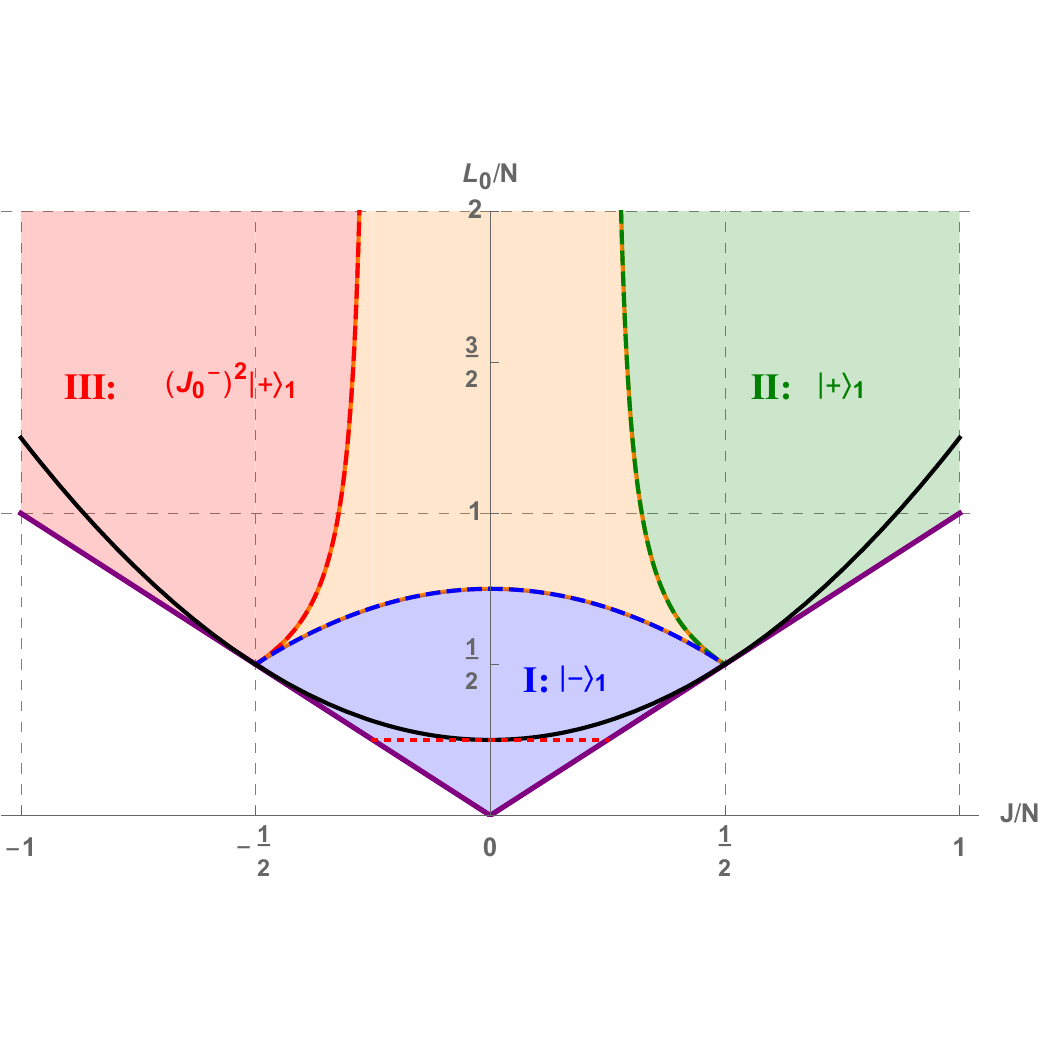} % Math'ca notes on 2019.05.14
  \vspace*{-7ex}
  \caption{\sl The phase diagram on the $J$-$L_0$ plane in the NS
  sector.  See the caption of Fig.~\ref{fig:regions} for explanation of
  regions and curves.  Because of the flip involved in the map
  \eqref{spfl+flip}, the positions of regions II and III are switched
  relative to the R sector (see Fig.~\ref{fig:regions}).
  \label{fig:regions_NS}}.
\end{center}
  \end{quote} 
\end{figure}

If nothing condenses, the entropy in the R sector is given by
\eqref{haxa8May19}, which becomes, in the NS sector,
\begin{align}
 S_0=4\left[c\,C \left({N^2\over 4} - J^2\right) \left(L_0 - {N\over 2}\right)\right]^{1/4}.
\end{align}

Region I, which was defined for the R sector by \eqref{regionI}, is
defined in the NS sector by
\begin{align}
 |J|\le N_p\le {N\over 2} - {J^2\over N}
\end{align}
and the entropy there is
\begin{align}
S_{\rm I}=4\left[{2c\,C\over 27} \Bigl(L_0 (L_0^2 - 9 J^2) + (L_0^2 + 3 J^2)^{3/2}\Bigr)\right]^{1/4}.
\end{align}
In region I, what condenses is $\ket{-}_1$, but this is nothing but the
vacuum (of a single copy of $M$).  So, in the NS sector, it is more
appropriate to say that, in this region, the excitations have not yet
fully occupied the $N$ copies.

Regions II and III, which were given for the R sector by
\eqref{regionII} and \eqref{regionIII}, are defined in the NS sector by
\begin{align}
N_p\ge {N^2 \mp 8 N J + 4 J^2 \over 4 (N \mp 4 J)},\qquad
 N_p\ge |J|,\qquad |J|\le N,
\end{align}
where the $-$ ($+$) sign is for region II (III)\@.
The entropy is
\begin{align}
S_{\rm II,III} 
&=4\biggl[{2c\,C\over 243}
 \Bigl((4N \mp J - 3 L_0) (-N \mp 2 J + 3 L_0) (2N \mp 5 J + 3 L_0) 
 \notag\\[-2ex]
 &\hspace{15ex}
 + (4N^2 + 7 J^2 - 6N L_0 + 9 L_0^2 \mp 2 J (N + 6 L_0))^{3/2}\Bigr)\biggr]^{1/4}.
\end{align}

See Fig.~\ref{fig:regions_NS} for the phase diagram on the $J$-$L_0$
plane in the NS sector.

If we take into account of the spectral flows, only the states above in
the range $J\in [-{1\over 2}N,{1\over 2}N]$ are dominant.  In the range
$J\in [(\eta - {1\over 2})N,(\eta + {1\over 2})N]$, $\eta\in\bbZ$, the
spectral-flowed states by the parameter $\eta$ are dominant.

%%%%%%%%%%%%%%%%%%%%%%%%%%%%%%%%%%%%%%%%%%%
%%%        MAIN TEXT ENDS HERE
%%%%%%%%%%%%%%%%%%%%%%%%%%%%%%%%%%%%%%%%%%%

\end{document}